\documentclass[%
 reprint,
superscriptaddress,
 amsmath,amssymb,
 aps,
pra,
]{revtex4-2}
\usepackage{graphicx,dcolumn,bm,physics}
\usepackage[colorlinks=true,allcolors=blue]{hyperref}
\begin{document}

\title{Continuous-wave quantum light control via engineered Rydberg-induced dephasing}

\author{Iason Tsiamis}
\email{iason.tsiamis@fu-berlin.de}
\affiliation{Dahlem Center for Complex Quantum Systems and Fachbereich Physik, Freie Universit\"{a}t Berlin, 14195 Berlin, Germany}
\affiliation{The Niels Bohr Institute, University of Copenhagen, Blegdamsvej 17, DK-2100 Copenhagen, Denmark}

\author{Oleksandr Kyriienko}
\affiliation{Department of Physics and Astronomy, University of Exeter, Stocker Road, Exeter EX4 4QL, United Kingdom}

\author{Anders S. S{\o}rensen}
\affiliation{Center for Hybrid Quantum Networks (Hy-Q), The Niels Bohr Institute, University of Copenhagen, DK-2100 Copenhagen {\O}, Denmark}

\begin{abstract}
We analyze several implementations of all-optical single-photon transistors (SPTs) operating in the continuous-wave (cw) regime, as presented in the companion paper [\href{https://doi.org/10.1103/7q3h-nvm6}
{Phys.\ Rev.\ A~\textbf{113}, L011701 (2026)}]. The devices rely on ensembles of Rydberg atoms interacting via van der Waals interactions. Under electromagnetically induced transparency (EIT), a weak probe field is fully transmitted through the atomic ensemble in the absence of control photons. Exciting a collective Rydberg state with a single control photon breaks the EIT condition, thereby strongly suppressing the probe transmission. We show how collective Rydberg interactions in an atomic ensemble, confined either in an optical cavity or in free space, give rise to two distinct probe-induced dephasing mechanisms. These processes localize the control excitations, extend their lifetimes, and increase the device efficiency. We characterize the SPTs in terms of control-photon absorption probability and probe gain, supported by numerical simulations of realistic one- and three-dimensional ensembles. The proposed cw devices complement previously demonstrated SPTs and broaden the toolbox of quantum light manipulation circuitry.
\end{abstract}

\maketitle

\section{\label{sec:intro}Introduction}
The interaction between atoms and light has been a central research topic since the early days of quantum mechanics. In recent decades, significant progress has been made in realizing a quantum interface between a large number of atoms, known as an atomic ensemble, and quantum light. This has been achieved by effectively coupling a collective superposition state of many atoms to quantum light \cite{Hammerer2010}. The interaction of light with multiatom ensembles has resulted in numerous applications for quantum information processing, including quantum memories \cite{Jaksch2000, Lukin2001} and quantum repeaters \cite{Kuzmin2003, Chaneliere2005}. These applications take advantage of the collective quantum behavior of the atomic ensemble to store and process quantum information encoded in light, enabling the development of more advanced quantum technologies.

Atomic ensembles with Rydberg excitations have captured significant attention in recent years due to their characteristic non-linear properties arising from their strong van der Waals interaction. Their utilization has contributed to significant progress in quantum optics \cite{Browaeys2020,Adams2020,Henriet2020,Saffman2010}, enabling the realization of strong photon-photon interactions \cite{Firstenberg2013, Gorshkov2011}. This has led to proposals for numerous applications such as Wigner crystallization \cite{ Otterbach2013}, Rydberg atomic qubits \cite{Xu2021}, and quantum gates \cite{Das2016,Higgins2017,Levine2019,McDonnell2022,Bluvstein2022}. The high degree of control and strong interaction also makes Rydberg atoms a promising platform for quantum simulation \cite{Labuhn2016,Bernien2017,Sbroscia2020,Ebadi2021,Semeghini2021,Daley2022} and quantum optimization \cite{LeoZ2020,Weggemans2022,Ebadi2022b}. 

The potential of Rydberg atoms as a platform for efficient optical control has been demonstrated by several Rydberg-based coherent switches and transistors, offering quantum optical control of light by light \cite{Tiarks2014, Gorniaczyk2014, Baur2014, Gorniaczyk2016}. The fundamental limit of such a device is the single-photon transistor (SPT), which was first proposed for a surface plasmon coupled to an atom in a waveguide \cite{Chang2007}. Similar SPTs have been realized in various other systems \cite{Chen2013,Tiecke2014,Sun2018,Aghamalyan2019}. However, these devices operate in the regime where the control photon and probe are applied separately, preventing continuous-wave operation. A continuous-wave (cw) SPT was proposed in the microwave regime \cite{Kyriienko2016,Royer2018,Grimsmo2021,ZWang2022}, but here we propose a cw SPT in the optical regime.

In this paper, we examine several implementations of cw SPTs based on an ensemble of driven Rydberg atoms, presented in the companion paper \cite{letter}. The two versions we theoretically study place the ensemble in free space or inside a single-sided cavity. For the proposed devices, the absence or presence of a single control photon is distinguished by measuring the probe field. When there is no control photon, the system is in the electromagnetically induced transparency (EIT) regime, which allows the probe field to propagate through the ensemble without loss \cite{EITrev,Hammerer2010}. When a single control photon enters the system and is converted into a long-lived collective Rydberg excitation, it results in interaction-induced blockade \cite{Urban2009} and lifts the EIT condition of the probe field, which leads to a measurable signal.

To achieve a long storage time and high gain of the transistor, we engineer the probe-induced dephasing of the control photons. This enables cw operation of all driving fields throughout the entire protocol, thereby simplifying the protocol and extending the possible applications of the transistor. Notably, our proposed devices can serve as efficient optical single-photon detectors with a large signal-to-noise ratio \cite{ Kyriienko2016}. We further believe that this study opens new possibilities for the application of Rydberg atoms in quantum information processing and quantum communication.

The remainder of this paper is structured as follows: In Sec. \ref{sec:cav}, the realization of the proposed cw SPT device based on a single-sided cavity is discussed, while in Sec. \ref{sec:fs}, the free-space version of the device is analyzed. Both sections are divided into five subsections. In Secs. 
\ref{sec:modcav} and \ref{sec:sysfs} the model is introduced, while in Secs. \ref{sec:probcav} and \ref{sec:fsprob} the modification of the scattering properties of the probe field and the dephasing induced on the control branch are examined. In Secs. 
\ref{subsec:imcav} and \ref{sec:fsim}, the optimal impedance matching conditions for control photon storage are theoretically estimated. In Secs. \ref{sec:simucav} and \ref{sec:simfs}, the Monte Carlo wave-function method used to simulate the system is presented, and in Secs. \ref{sec:rescav} and \ref{sec:resfs}, the results of the simulation are used to characterize the efficiency of the proposed devices and are compared with the theoretical estimates. Finally, in Sec. \ref{sec:concl}, we provide the conclusions of our work.

\section{\label{sec:cav}Cavity}

\subsection{\label{sec:modcav}Model}
\begin{figure}[t!]
\includegraphics[width=1.\columnwidth]{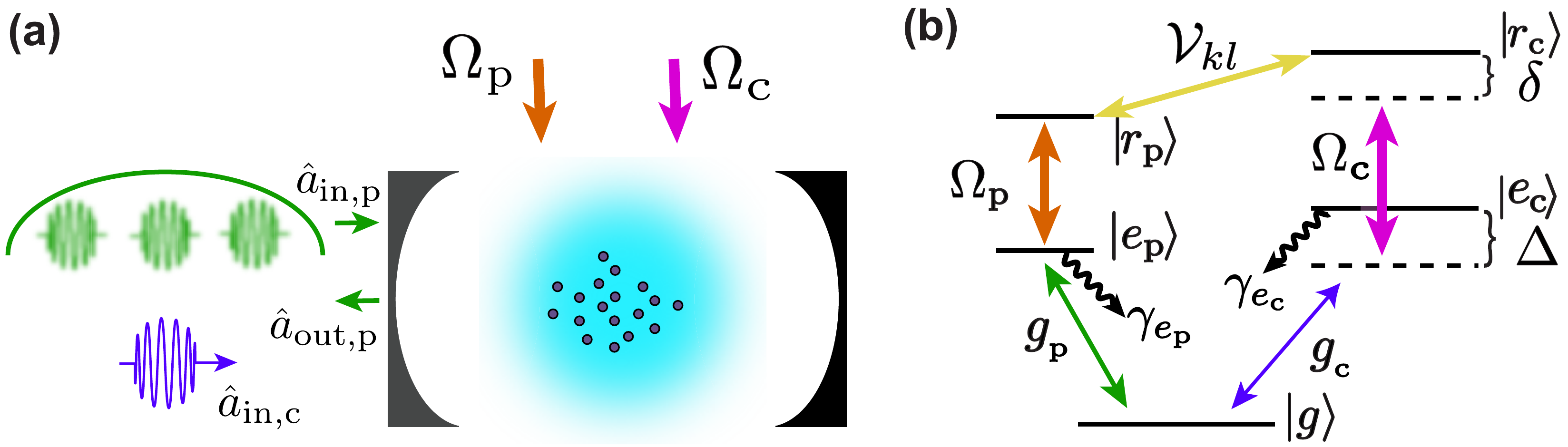}
\caption{(a) Sketch of the system, showing the cavity version of the Rydberg-based single-photon transistor. An ensemble of Rydberg atoms in an optical cavity is continuously probed by a weak coherent field with strength $|\alpha_{\mathrm{in,p}}|^2$, which is fully reflected under EIT conditions, while driven by two classical drives $\Omega_\mathrm{c}$ and $\Omega_\mathrm{p}$. Upon sending a control photon $\hat{a}_{\mathrm{in,c}}$ to the cavity, it is converted into a Rydberg excitation, inducing an energy-level shift that modifies the reflection of the probe. (b) Atomic level scheme. The relevant atomic states form two $\Xi$ subsystems (branches): a probe branch ($\{ |g\rangle, |e_\mathrm{p}\rangle, |r_\mathrm{p}\rangle \}$), responsible for EIT-based probing by the input field $\alpha_{\mathrm{in,p}}$ under the drive $\Omega_\mathrm{p}$, and a control branch ($\{ |g\rangle, |e_\mathrm{c}\rangle, |r_\mathrm{c}\rangle \}$), responsible for converting a single-photon pulse into a stored Rydberg excitation $|r_\mathrm{c}\rangle$, which breaks the probe EIT via the interbranch van der Waals interaction $\mathcal{V}_{kl}$.}
\label{fig:sketch}
\end{figure}

We begin by introducing the version of the SPT consisting of a Rydberg atomic cloud placed in a single-sided cavity. The cavity is subject to two driving fields and two weak fields, serving as probe and control fields respectively, as illustrated in Fig.~\ref{fig:sketch}(a). In Fig.~\ref{fig:sketch}(b) we show the relevant energetic structure of the Rydberg atoms, which was previously used for pulsed operation \cite{Hao2019}. The Hamiltonian for the system is composed of several parts
\begin{align}
\label{eq:H}
&\hat{\mathcal{H}} = \hat{\mathcal{H}}_{\mathrm{probe}} + \hat{\mathcal{H}}_{\mathrm{control}} + \hat{\mathcal{H}}_{\mathrm{Ryd}} + \hat{\mathcal{H}}_{\mathrm{input}} + \hat{\mathcal{H}}_{\mathrm{res}}.
\end{align}
Here, $\hat{\mathcal{H}}_{\mathrm{probe}}$ represents the atomic branch used for probing (left branch in Fig.~\ref{fig:sketch}(b)), $\hat{\mathcal{H}}_{\mathrm{control}}$ corresponds to the branch responsible for single-photon storage of the control field [right branch in Fig.~\ref{fig:sketch}(b)], and $\hat{\mathcal{H}}_{\mathrm{Ryd}}$ describes the Rydberg interaction between the atoms. Additionally, we introduced the coupling to a continuum $\hat{\mathcal{H}}_{\mathrm{input}}$, providing input and output for the cavity, and the reservoir Hamiltonian $\hat{\mathcal{H}}_{\mathrm{res}}$.

The probe part of the Hamiltonian is given by
\begin{align}
\label{eq:H_probe}
\hat{\mathcal{H}}_{\mathrm{probe}}=-\sum\limits_{l=1}^{N}\hbar (g_{\mathrm{p}} \hat{a}_{\mathrm{p}} \hat{\sigma}_{e_\mathrm{p}g}^l+\Omega_\mathrm{p}\hat{\sigma}_{r_\mathrm{p}e_\mathrm{p}}^l+ \mathrm{H.c.}) ,
\end{align}
where the creation (annihilation) operator $\hat{a}^{\dagger}_{\mathrm{p}}$ ($\hat{a}_\mathrm{p}$) describes a probe cavity photon and $g_{\mathrm{p}}$ is the coupling constant between atoms and probe cavity photons. The energy levels $|e_ \mathrm{p}\rangle$ and $|r_ \mathrm{p}\rangle$ are coupled by a classical drive with Rabi frequency of $\Omega_\mathrm{p}$. The transition of the $l$th atom between states $\ket{m}$ and $\ket{n}$ is described by the operator $\hat{\sigma}_{mn}^l = \ket{m_l} \bra{n_l}$, where ${m,n}\in\{{g,e_\mathrm{p},r_\mathrm{p},e_\mathrm{c},r_\mathrm{c}}\}$, and $N$ is the total number of atoms comprising the ensemble. The rotating wave approximation has been performed and the excitation scheme corresponds to EIT conditions, where all fields are resonant with the corresponding transitions.

The Hamiltonian that describes the control branch reads 
\begin{align}
\notag
\hat{\mathcal{H}}_{\mathrm{control}}=&\sum\limits_{l=1}^{N}\hbar[ \Delta\hat{\sigma}_{e_\mathrm{c}e_\mathrm{c}}^l +\delta\hat{\sigma}_{r_\mathrm{c}r_\mathrm{c}}^l - (\Omega_\mathrm{c}\hat{\sigma}_{r_\mathrm{c}e_\mathrm{c}}^l +\Omega_\mathrm{c}^*\hat{\sigma}_{e_\mathrm{c}r_\mathrm{c}}^l ) \\ &+ (g_\mathrm{c} \hat{a}_\mathrm{c}\hat{\sigma}_{e_\mathrm{c}g}^l +g_\mathrm{c}^*\hat{a}^{\dagger}_{\mathrm{c}} \hat{\sigma}_{ge_\mathrm{c}}^l )  ],
\label{eq:H_control}
\end{align}
where we have used a rotating frame with respect to the classical drive frequency $\omega_{2,\mathrm{c}}$ and the resonance frequency of the cavity $\omega_{1,\mathrm{c}}$. The detuning $
\Delta$ is defined as $\Delta = \omega_{e_\mathrm{c}g} - \omega_{1,\mathrm{c}}$, while the two photon detuning is $\delta = \Delta - (-\omega_{r_\mathrm{c}e_\mathrm{c}} + \omega_{2,\mathrm{c}})$. Both are depicted in the energy diagram of Fig.~\ref{fig:sketch}(b). The atomic transition frequencies are $\omega_{e_\mathrm{c}g}=\omega_{e_\mathrm{c}}-\omega_g$ and $\omega_{r_\mathrm{c}e_\mathrm{c}}=\omega_{r_\mathrm{c}}-\omega_{e_\mathrm{c}}$, where the frequencies $\omega_g$, $\omega_{e_\mathrm{c}}$, $\omega_{r_\mathrm{c}}$ correspond to the energies of states $\ket{g}$, $\ket{e_\mathrm{c}}$, $\ket{r_\mathrm{c}}$, respectively. The operator $\hat{a}^{\dagger}_\mathrm{c}$ describes the control photon, which interacts with each atom with coupling strength $g_\mathrm{c}$, and the classical drive for the transition of states $\ket{e_\mathrm{c}}$ and $\ket{r_\mathrm{c}}$ is characterized by the Rabi frequency $\Omega_\mathrm{c}$.

The atoms in highly-excited Rydberg states $\ket{r_\mathrm{p}}$ and $\ket{r_\mathrm{c}}$ interact through the van der Waals interaction, which is modeled by the interaction Hamiltonian
\begin{equation}
\label{eq:H_Ryd}
\hat{\mathcal{H}}_{\mathrm{Ryd}} = \hbar \sum\limits^N_{\substack{l=1\\l\neq k}}\sum\limits^N_{k=1} \mathcal{V}_{kl} \hat{\sigma}_{r_\mathrm{p}r_\mathrm{p}}^l \otimes \hat{\sigma}_{r_\mathrm{c}r_\mathrm{c}}^k,
\end{equation}
where $\mathcal{V}_{kl}=C_6/\rho_{kl}^6$, with $\rho_{kl}$ denoting the distance between atoms and $C_6$ is an interaction coefficient \cite{Saffman2010}. 
We analyze the system's capability to store control photons and manipulate the probe by distinguishing between the control and probe branches. Our analysis assumes that the control branch receives a single control photon, whereas the probe branch can receive multiple probe photons. The Hamiltonian (\ref{eq:H_Ryd}) only considers interaction between atoms excited in the Rydberg states of different branches. This approach is justified in two cases: First, when the principal quantum number of the Rydberg state in the control branch significantly exceeds that of the Rydberg state in the probe branch, resulting in much larger dipole matrix elements between $\ket{r_\mathrm{c}}$ and its neighboring states compared with those of $\ket{r_\mathrm{p}}$ and its neighboring states. Second, if the Rydberg states $\ket{r_\mathrm{c}}$ and $\ket{r_\mathrm{p}}$ are close to a Förster resonance \cite{Saffman2008}, leading to a near-zero energy difference between the combined energies of states $\ket{r_\mathrm{c}}$ and $\ket{r_\mathrm{p}}$ and those of a pair of neighboring states. Both cases are thoroughly discussed in Appendix \ref{appendix:a}. Finally, in order to model the input-output of the system, we add the coupling to the environment and reservoir Hamiltonians $\hat{\mathcal{H}}_{\mathrm{input}}+\hat{\mathcal{H}}_{\mathrm{res}}$ in the standard form \cite{Clerk2010, WallsMilburn}.

Using the Hamiltonian \eqref{eq:H}, we can derive the equations of motion for the operators of the system. Using these equations, we can quantify the storage capability of the control branch and the change in the transmission properties of the probe branch \cite{Witthaut2010,Fan2010,Dalibard1992}. We assume that the probe branch operates on a much faster timescale than the control branch, indicating a substantial difference in the EIT-based bandwidth of the probe and the Raman-based bandwidth of the control branch. This enables us to examine the probe and control branches separately in the two following sections.


\subsection{\label{sec:probcav}Probe-induced dephasing}
Initially, let us focus on the branch where the continuous probe field is incident [left branch depicted in Fig.~\ref{fig:sketch}(b)]. In this section, we investigate the changes in the scattering properties of the probe field based on the presence or absence of a control photon, which is the primary role of the SPT \cite{Chang2007}. Additionally, we examine the effective dephasing mechanisms induced by the continuous probe field on the stored control excitation through the Rydberg interaction.

The dynamics of the probe branch can be fully described by the equations of motion (EOMs) derived from the Hamiltonians $\hat{\mathcal{H}}_\mathrm{probe}+\hat{\mathcal{H}}_\mathrm{Ryd}$ and the Lindblad jump operators 
\begin{align}\label{eq:kjum}
&\hat{L}_{\kappa_\mathrm{p}}=\sqrt{2\kappa_\mathrm{p}}\hat{a}_\mathrm{p},\\ 
&\hat{L}^l_{ge_\mathrm{p}}=\sqrt{2\gamma_{e_\mathrm{p}}}\hat{\sigma}^l_{ge_\mathrm{p}}. \label{eq:gjum}
\end{align}
These operators account for the losses to the environment of the intracavity probe photons and of the $l$th atom in the probe excited state $\ket{e_\mathrm{p}}$ through spontaneous emission, respectively. We further introduce the decay rate of the cavity probe photons $\kappa_\mathrm{p}$ and the decay rate of the probe excited state $\gamma_{e_\mathrm{p}}$. By considering independent Lindblad operators for each atom, we implicitly assume that they are coupled to independent reservoirs. This implies that emitted photons cannot be absorbed by a different atom, which holds true for a low-density atomic ensemble within a cavity, where the optical depth of the ensemble is small, thereby reducing the probability of photon reabsorption by neighboring atoms.
\par The Heisenberg EOM is used to describe the evolution of system operators through  \begin{equation}\label{eq:heom}
\dot{\hat{o}}=\frac{i}{\hbar}[\hat{\mathcal{H},}\hat{o}]+\sum_j\left(\hat{L}^\dagger_j\hat{o}\hat{L}_j-\frac{1}{2}\{\hat{L}^\dagger_j\hat{L}_j,\hat{o}\}\right),
\end{equation}
where $\hat{o}$ can be any system operator, and $\hat{L}_j$ refers to the system's Lindblad operators. We note that for simplicity we have omitted the Langevin noise operators associated with the decays. This is due to the fact that we consider the decay of optical excitation where the reservoir is typically in the vacuum state. The corresponding noise operators therefore will always have expectation values corresponding to vacuum noise. Since vacuum noise will never give rise to any excitations of the system, so the latter can be neglected in the following. The resulting EOMs for the operators associated with the probe branch are found to be
\begin{align}
\label{eq:a_cp}
&\dot{\hat{a}}_\mathrm{p} = -\kappa_\mathrm{p}\hat{a}_\mathrm{p} + \sqrt{2 \kappa_\mathrm{p}} \hat{a}_\mathrm{in,p} + ig^*_\mathrm{p} \sum_{l=1}^N \hat{\sigma}_{ge_\mathrm{p}}^l,\\
\label{eq:sigma_ge_cp}
&\dot{\hat{\sigma}}_{ge_\mathrm{p}}^l = -\gamma_{e_\mathrm{p}}  \hat{\sigma}_{ge_\mathrm{p}}^l + i g_\mathrm{p} \hat{a}_\mathrm{p} + i \Omega_\mathrm{p} \hat{\sigma}_{gr_\mathrm{p}}^l ,\\
&
\dot{\hat{\sigma}}_{g{r_\mathrm{p}}}^l=-i\sum_{\substack{k=1\\k\neq l}}^N\mathcal{V}_{kl}\hat{\sigma}^k_{r_\mathrm{c}r_\mathrm{c}}\otimes\hat{\sigma}_{gr_\mathrm{p}}^l+i\Omega_\mathrm{p}^*\hat{\sigma}_{ge_\mathrm{p}}^l.
\label{eq:sigma_gr_cp}
\end{align}
Equation \eqref{eq:a_cp} describes the probe cavity photon's dynamics, accounting for losses to the environment and the input probe field. Equation \eqref{eq:sigma_ge_cp} captures the evolution of the probe excited state $\ket{e_\mathrm{p}}$ of the $l$th atom and the interaction with the cavity field and the excitation to the Rydberg state. Equation \eqref{eq:sigma_gr_cp} tracks the evolution of the $l$th atom's Rydberg excitation of the probe branch, which interacts with the Rydberg excitations of the control branch $\ket{r_\mathrm{c}}$ of other atoms through van der Waals interactions. We have approximated that the number of atoms is sufficiently large, such that we are operating below the saturation point so that the operators $\hat{\sigma}_{mm}$, where $m\in\{g,e_{\mathrm{p}},r_{\mathrm{p}}\}$, can be neglected.

The incident probe field is subject to the cavity boundary condition, described by the input-output relation \cite{Gardiner,Molmer1993,Murray2016} for the probe cavity photons 
\begin{equation}    \hat{a}_\mathrm{in,p}+\hat{a}_\mathrm{out,p}=\sqrt{2\kappa_\mathrm{p}}\hat{a}_\mathrm{p}.
    \label{eq:input_cp}
\end{equation}
The probe input field is considered to be prepared in a coherent state with classical amplitude $\alpha_\mathrm{in,p}$ and vacuum fluctuations $\delta\hat{a}_\mathrm{in,p}$, i.e., $\hat{a}_\mathrm{in,p}=\alpha_\mathrm{in,p}+\delta\hat{a}_\mathrm{in,p}$. 
\par We initially describe the system in the absence of any stored excitation in the control branch, which serves as a reference. By subtracting this reference, we obtain a purely Rydberg-interaction related description. When there is no stored excitation in the control branch,  Eq.~(\ref{eq:sigma_gr_cp}) simplifies to
\begin{equation}
\dot{\hat{\sigma}}_{g{r_\mathrm{p}}}^l=i\Omega_\mathrm{p}^*\hat{\sigma}_{ge_\mathrm{p}}^l.
 \label{eq:sigma_grn_cp}
\end{equation}
Moving to the frequency domain under the Fourier transform
\begin{equation}
\hat{o}[\omega]=\int dt e^{i\omega t}\hat{o}[t],
 \label{eq:four}
\end{equation}
where $\omega$ is defined with respect to the resonance frequency of the cavity, the solutions of the EOMs (\ref{eq:a_cp}), (\ref{eq:sigma_ge_cp}), and (\ref{eq:sigma_grn_cp}) are obtained. The solutions have the form $\hat{o}=\overline{o}+\mathcal{O}(\delta\hat{a}_\mathrm{in,p})$ consisting of a classical part proportional to the coherent probe amplitude $\alpha_\mathrm{in,p}$ and a part proportional to the quantum fluctuations $\delta\hat{a}_\mathrm{in,p}$. The classical parts of the solutions in the absence of stored control excitation are
\begin{align}
\label{eq:ac_cp}
&\overline{a}_\mathrm{p}[\omega]=\frac{\sqrt{2\kappa_\mathrm{p}}\alpha_\mathrm{in,p}[\omega]}{\kappa_\mathrm{p}-i\omega+\frac{\omega |g_\mathrm{p}|^2N}{\omega(\gamma_{e_\mathrm{p}}-i\omega)+i|\Omega_\mathrm{p}|^2}},\\
&\label{eq:sigmac_ge_cp}
\overline{\sigma}_{ge_\mathrm{p}}[\omega]=\frac{ g_\mathrm{p}\omega \sqrt{2\kappa_\mathrm{p}}\alpha_\mathrm{in,p}[\omega]}{(\kappa_\mathrm{p}-i\omega)(\omega(\gamma_{e_\mathrm{p}}-i\omega)+i|\Omega_\mathrm{p}|^2)+\omega|g_\mathrm{p}|^2N},
\\&\label{eq:sigmac_gr_cp}
\overline{\sigma}_{gr_\mathrm{p}}[\omega]=\frac{-i\Omega_\mathrm{p} g_\mathrm{p} \sqrt{2\kappa_\mathrm{p}}\alpha_\mathrm{in,p}[\omega]}{(\kappa_\mathrm{p}-i\omega)(\omega(\gamma_{e_\mathrm{p}}-i\omega)+i|\Omega_\mathrm{p}|^2)+\omega|g_\mathrm{p}|^2N}.
\end{align}

Equipped with these solutions, we can evaluate the scattering behavior of the system. 
The measure of the scattering properties is the reflection coefficient, described by the relation $\hat{a}_\mathrm{out,p}[\omega]=R_\mathrm{p}[\omega]\hat{a}_\mathrm{in,p}[\omega]$. Using the input-output condition (\ref{eq:input_cp}) and the solution of the intracavity probe field (\ref{eq:ac_cp}), the reflection coefficient in the absence of a control excitation is found to be 
\begin{equation}\label{eq:ref}
R_\mathrm{p}[\omega]=\frac{2\kappa_\mathrm{p}}{\kappa_\mathrm{p}-i\omega+\frac{g_\mathrm{p}^2N\omega}{\omega(\gamma_{e_\mathrm{p}}-i\omega)+|\Omega_\mathrm{p}|^2}}-1.
\end{equation}
We note that for the probe field being resonant with the cavity, i.e., $\omega=0$, the probe field is fully reflected $R_\mathrm{p}[\omega=0]=1$. This is due to the device operating under EIT conditions.
\par Subsequently, we turn to the description of the system in the presence of a stored control excitation. To achieve a description of solely the Rydberg-associated processes, the solutions of the probe branch's EOMs in the absence of any stored control excitation (\ref{eq:ac_cp})-(\ref{eq:sigmac_gr_cp}) are used as a reference frame. In that spirit, a new set of shifted operators is introduced by subtracting the reference frame from the original operators. The shifted operators of the probe branch are
\begin{align}
&\delta\hat{a}_\mathrm{p}[\omega]=\hat{a}_\mathrm{p}[\omega]-\overline{a}_\mathrm{p}[\omega],\label{eq:da}
\\
&
\delta\hat{\sigma}_{{ge}_\mathrm{p}}^l[\omega]=\hat{\sigma}_{{ge}_\mathrm{p}}^l[\omega]-\overline{\sigma}_{{ge}_\mathrm{p}}[\omega], \label{eq:dse}
\\
&
\delta\hat{\sigma}_{{gr}_\mathrm{p}}^l[\omega]=\hat{\sigma}_{gr_\mathrm{p}}^l[\omega]-\overline{\sigma}_{gr_\mathrm{p}}[\omega], \label{eq:dsr}
\\
&\label{eq:dain}
\delta\hat{a}_\mathrm{{in,p}}[\omega]=\hat{a}_\mathrm{in,p}[\omega]-\alpha_\mathrm{in,p}[\omega].
\end{align}
The EOMs of the shifted operators are obtained by substituting the definitions of the shifted operators (\ref{eq:da})-(\ref{eq:dsr}) into the probe branch's EOMs (\ref{eq:a_cp})-(\ref{eq:sigma_gr_cp}), which includes the potential presence of a Rydberg excitation on the control branch. The set of EOMs for the shifted probe branch operators in the frequency domain reads
\begin{align}
&-i\omega\delta\hat{a}_\mathrm{p}=-\kappa_\mathrm{p} \delta\hat{a}_\mathrm{p}+\sqrt{2\kappa_\mathrm{p}}\delta\hat{a}_\mathrm{in,p}
+ig^*_\mathrm{p}\sum_{l=1}^N\delta\hat{\sigma}_{ge_\mathrm{p}}^l,\label{eq:shifta}
\\&
-i\omega\delta\hat{\sigma}_{ge_\mathrm{p}}^l=-\gamma_{e_\mathrm{p}}\delta\hat{\sigma}_{ge_\mathrm{p}}^l+ig_\mathrm{p}\delta\hat{a}_\mathrm{p}+i\Omega_\mathrm{p}\delta\hat{\sigma}^l_{gr_\mathrm{p}},\label{eq:shifts}
\\&
-i\omega\delta\hat{\sigma}_{gr_\mathrm{p}}^l=-i\sum_{\substack{k=1\\k\neq l}}^N\mathcal{V}_{kl}\hat{\sigma}_{r_\mathrm{c}r_\mathrm{c}}^k\otimes\delta\hat{\sigma}_{gr_\mathrm{p}}^l+i\Omega_\mathrm{p}^*\delta\hat{\sigma}_{ge_\mathrm{p}}^l \nonumber
\\&\qquad\qquad\qquad+\frac{i\sqrt{2\kappa_\mathrm{p}}g_\mathrm{p}\alpha_\mathrm{in,p}}{\Omega_\mathrm{p}(\kappa_\mathrm{p}-i\omega)}\sum_{\substack{k=1\\k\neq l}}^N\mathcal{V}_{kl}\hat{\sigma}_{r_\mathrm{c}r_\mathrm{c}}^k.\label{eq:shift}
\end{align}
As expected a feeding term appears in Eq. (\ref{eq:shift}) in the presence of a stored excitation on the control branch, resulting from the Rydberg-Rydberg interaction between excitations in the probe and control branches.
This originates from the fact that in the shifted frame the EOMs of the probe branch describe solely the Rydberg-associated processes.
\par The solutions of these EOMs are of the form $\hat{o}[\omega]=\overline{o}[\omega]\hat{\sigma}_{r_\mathrm{c}r_\mathrm{c}}+\mathcal{O}(\delta\hat{a}_\mathrm{in}[\omega])$, where the first term is a classical term proportional to the amplitude of the probe field $\alpha_\mathrm{in}$, conditioned on the presence of a stored control excitation. The second term is a weak quantum term proportional to the fluctuation operator $\delta\hat{a}_\mathrm{in}$. In the shifted frame this operator acts on the vacuum state. Hence, for the normally ordered products that appear below, we can simply ignore this operator. By considering the probe input field resonant with the cavity, i.e., $\omega=0$, which corresponds to a field constant in time, the solutions for the probe cavity field and the probe excited state transition operator are found to be

\begin{align}
&\delta\hat{a}_\mathrm{p}[\omega=0]=-\sum_{k=1}^N\sqrt{\frac{2}{\kappa_\mathrm{p}}}\frac{\alpha_\mathrm{in,p}C_\mathrm{b,p}^k}{1+C_\mathrm{b,p}^k}\hat{\sigma}_{r_\mathrm{c}r_\mathrm{c}}^k,\label{eq:da1}\\
&
\delta\hat{\sigma}_\mathrm{ge_\mathrm{p}}^l[\omega=0]=\sum_{k=1}^N\frac{ig_\mathrm{p}\sqrt{\frac{2}{\kappa_\mathrm{p}}}}{\gamma_{e_\mathrm{p}}
+\frac{|\Omega_\mathrm{p}|^2}{i \mathcal{V}_{kl}}}\frac{\alpha_\mathrm{in,p}}{1+C_\mathrm{b,p}^k}\hat{\sigma}^k_{r_\mathrm{c}r_\mathrm{c}}.\label{eq:ds1}
\end{align}
The blockaded cooperativity for the probe branch due to a stored excitation in the control Rydberg state of the $k$th atom is here introduced as $C_\mathrm{b,p}^k=\sum_{\substack{l=1\\l\neq k}}^NC_\mathrm{b1,p}^{k,l}$, being the cooperativity of the atoms included in the Rydberg blockaded sphere around the $k$th atom. The single atom blockaded cooperativity of the $l$th atom due to the $k$th atom being excited in the Rydberg state of the control branch is given by $C_\mathrm{b1,p}^{k,l}=(|g_\mathrm{p}|^2/\kappa_\mathrm{p})/(\gamma_{e_\mathrm{p}}-i|\Omega_\mathrm{p}|^2/\mathcal{V}_{kl})$. For fully blockaded atoms $\mathcal{V}_{kl}\gg |\Omega_\mathrm{p}|^2/\gamma_{e_\mathrm{p}}$ this yields the standard expression for cooperativity of the probe branch $C_\mathrm{p}=|g_\mathrm{p}|^2N/(\gamma_{e_\mathrm{p}}\kappa_\mathrm{p})$,  whereas outside the blockade regime $\mathcal{V}_{kl}\ll |\Omega_\mathrm{p}|^2/\gamma_{e_\mathrm{p}}$, the blockaded cooperativity is suppressed. 
\par At this point the main function of the SPT becomes evident: in the presence of a stored control excitation in the Rydberg state of the $k$th atom $\ket{r_\mathrm{c}^k}$, the scattering properties of the probe are altered from the full reflection observed in Eq. (\ref{eq:ref}). This is confirmed by using the input-output relation (\ref{eq:input_cp}), the solution for the shifted field (\ref{eq:da1}), and the mapping between the shifted and the initial operators (\ref{eq:da}). From these expressions the reflection coefficient on resonance is found to be  
\begin{equation}\label{eq:refk}
R_\mathrm{p}^k[\omega=0]=\frac{1-C_\mathrm{b,p}^k}{1+C_\mathrm{b,p}^k},
\end{equation}
depending on the atom where the control photon is stored. To gain some intuition about this coefficient, let us examine the reflection coefficient in the absence of the driving field $\Omega_\mathrm{p}$, which decouples $\ket{r_\mathrm{p}}$ from the rest of the system, effectively reducing the probe branch to a 2-level system. In this case, using the reflection coefficient in  Eq. (\ref{eq:ref}) is $R_\mathrm{p}[\omega=0]=(1-C_\mathrm{p})/(1+C_\mathrm{p})$. By analogy, Eq. (\ref{eq:refk}) can be interpreted as arising from an effective two-level system within a sphere centered at the excited $k$th atom with a radius equal to the Rydberg blockade radius. 

\par As shown in Eq. (\ref{eq:refk}), in the presence of a stored control excitation the blockaded cooperativity $C_\mathrm{b,p}^k$ modifies the scattering properties of the system. This modification leads in turn to the extraction of information regarding the presence and/or location of the stored control excitation. The extraction of information about the presence or location of the stored excitation in the control branch leads to dephasing processes in that branch. These processes arise from the two Lindblad operators (\ref{eq:kjum}) and (\ref{eq:gjum}), which in the shifted frame read
\begin{align}\label{eq:kjump}
&\hat{L}_{\kappa_\mathrm{p}}=-\sum_{k=1}^N\frac{\alpha_\mathrm{in,p}C_\mathrm{b,p}^k}{1+C_\mathrm{b,p}^k}\hat{\sigma}_{r_\mathrm{c}r_\mathrm{c}}^k\equiv \sum_{k=1}^N\sqrt{2\gamma_{\kappa_\mathrm{p}}^k}\hat{\sigma}_{r_\mathrm{c}r_\mathrm{c}}^k,\\
&\hat{L}_{ge_\mathrm{p}}^l=\sum_{\substack{k=1\\k\neq l}}^N\frac{i2g_\mathrm{p}\sqrt{\frac{\gamma_{e_\mathrm{p}}}{\kappa_\mathrm{p}}}\alpha_\mathrm{in,p}}{\gamma_{e_\mathrm{p}}+\frac{|\Omega_\mathrm{p}|^2}{i \mathcal{V}_{kl}}}\frac{\hat{\sigma}^k_{r_\mathrm{c}r_\mathrm{c}}}{1+C_\mathrm{b,p}^k}\equiv \sum_{\substack{k=1\\k\neq l}}^N\sqrt{2\gamma_{ge_\mathrm{p}}^{k,l}}\hat{\sigma}_{r_\mathrm{c}r_\mathrm{c}}^k,\label{eq:gjump}
\end{align}
representing the dephasing of the stored control excitation from a change in cavity reflection of the probe and the dephasing due to spontaneous emission of the $l$th atom from the probe excited state $\ket{e_\mathrm{p}}$ respectively. The dephasing rates on the $k$th atom, associated with the two processes, are given by the relation $\gamma_i^k=\bra{r^k_\mathrm{c}}\frac{1}{2}\hat{L}_{i}^\dagger\hat{L}_{i}\ket{r^k_\mathrm{c}}/\braket{r_\mathrm{c}^k}$
\begin{align}
&\gamma^k_{\kappa_\mathrm{p}}=\sum_{k=1}^N\left|\gamma_{\kappa_\mathrm{p}}^k\right|^2=\sum_{k=1}^N\frac{2|C_\mathrm{b,p}^k|^2}{|1+C_\mathrm{b,p}^k|^2}|\alpha_\mathrm{in,p}|^2,\label{eq:krate}
\\&\gamma_{ge_\mathrm{p}}^{k,l}=\sum^N_{\substack{k=1\\k\neq l}}\left|\gamma_{e_\mathrm{p}}^{k,l}\right|^2=\sum_{\substack{k=1\\k\neq l}}^N\frac{\mathrm{Re}(C_\mathrm{b1,p}^{k,l})|\alpha_\mathrm{in,p}|^2}{|1+C_\mathrm{b,p}^k|^2}.\label{eq:grate}
\end{align}
\par Let us first focus on the dephasing operator $\hat{L}_{ge_\mathrm{p}}^l$ due to spontaneous emission of the $l$th atom from the probe excited state $\ket{e_\mathrm{p}}$. Its dependence on which atom decayed ($l$), provides information to the environment regarding the position of the collective Rydberg excitation leading to localization of the stored control excitation in the vicinity of $l$. This is evident by the presence of $\mathcal{V}_{kl}$ in the denominator, which localizes the stored excitation around the decayed atom, since $\gamma_{e_\mathrm{p}}^{k,l}$ is nonzero only for the atoms $k$ for which the distance $\rho_{kl}$ is shorter than the Rydberg blockade radius. 
\par The second dephasing operator $\hat{L}_{\kappa_\mathrm{p}}$ acts on the stored excitation due to a change in cavity reflection. This dephasing process does not reveal any information regarding the location of the stored excitation and, unlike $\hat{L}_{ge_\mathrm{p}}^l$, does not lead to localization. Instead, it reveals information only about the presence of the stored excitation, via the change of the reflected probe field, which is the mechanism behind the proposed SPT. Although this dephasing process is suppressed for $C_\mathrm{b,p}\ll 1$ as seen in Eq. (\ref{eq:krate}), for larger values of the blockaded cooperativity it is present and detectable as seen by the change in the reflection coefficient in Eq. (\ref{eq:refk}) compared with the full reflection observed in the absence of any stored excitation.  
\par Both dephasing processes are an intricate part of the functioning of the proposed device. $\hat{L}_{\kappa_\mathrm{p}}$ is the detectable process that is essential for the efficiency of the proposed device, but the $\hat{L}_{ge_\mathrm{p}}^l$ process leads to localization which provides a long lifetime for the stored excitation (see Sec. \ref{sec:rescav}). An important aspect of the current work is the balancing of these two processes for the optimization of the functionality of the SPT. 
\par Finally, we introduce the total collective dephasing rate of the control stored Rydberg excitation, due to the Rydberg-mediated processes of the probe branch, 
\begin{align}\label{eq:rater}
\gamma_{r}=\frac{\gamma_{\kappa_\mathrm{p}}+\sum_{l=1}^N\gamma_{ge_\mathrm{p}}^l}{N}=\sum_{k=1}^N\frac{2\left(|C_\mathrm{b,p}^k|^2+\mathrm{Re}(C_\mathrm{b,p}^k)\right)}{N|1+C_\mathrm{b,p}^k|^2}|\alpha_\mathrm{in,p}|^2.
\end{align}
This rate is proportional to the strength of the probe field $|\alpha_\mathrm{in,p}|^2$. We can thus adjust the probing strength $|\alpha_\mathrm{in,p}|^2$ to optimize impedance matching conditions for the control photon storage by counterbalancing decay processes on the control branch. This will be discussed in Sec. \ref{subsec:imcav}.


\subsection{\label{subsec:imcav}Impedance matching}
A critical step for the functioning of the SPT is the conversion of a control photon, incident to the cavity, to a Rydberg collective excitation with near unity probability. To achieve that, we analyze the scattering dynamics on the control branch and derive an analytical estimate for the impedance matching conditions.
\par The dynamics of the control branch can be fully described by the corresponding equations of motion derived from $\hat{H}_\mathrm{control}$ and the Lindblad jump operators 
\begin{align}
&\hat{L}_{\kappa_\mathrm{c}}=\sqrt{2\kappa_\mathrm{c}}\hat{a}_\mathrm{c},\\
&\hat{L}^l_{ge_\mathrm{c}}=\sqrt{2\gamma_{e_\mathrm{c}}}\hat{\sigma}^l_{ge_\mathrm{c}},\label{eq:gec}
\\&\hat{L}^l_{r_\mathrm{c}r_\mathrm{c}}=\sqrt{2\gamma_{r}}\hat{\sigma}^l_{r_\mathrm{c}r_\mathrm{c}}\label{eq:deph}.
\end{align}
 The first two Lindblad operators account for the losses to the environment of the intracavity control photon via the decay of the cavity and the decay of the $l$th atom from the control excited state $\ket{e_\mathrm{c}}$ via spontaneous emission, respectively. The third Lindblad operator (\ref{eq:deph}) accounts for the total effective dephasing of the control Rydberg state $\ket{r_\mathrm{c}}$ due to the decay dynamics of the probe branch's Lindblad operators (\ref{eq:kjum}) and (\ref{eq:gjum}) mediated by the Rydberg interaction, as derived in Sec. \ref{sec:probcav}. 
 \par The EOMs for the system operators associated with the control field are derived using the Heisenberg equation ($\ref{eq:heom}$) and read
\begin{align}
\label{eq:ac_dot}
&\dot{\hat{a}}_\mathrm{c} = -\kappa_\mathrm{c}\hat{a}_\mathrm{c} + \sqrt{2 \kappa_\mathrm{c}} \hat{a}_{\mathrm{in,c}} + ig^*_\mathrm{c} \sum_{l=1}^N\hat{\sigma}_{ge_\mathrm{c}}^l,\\
\label{eq:sigma_gec_dot}
&\dot{\hat{\sigma}}_{ge_{c}}^l = -(\gamma_{e_\mathrm{c}} + i \Delta) \hat{\sigma}_{ge_{c}}^l + i g_\mathrm{c} \hat{a}_\mathrm{c} + i \Omega_\mathrm{c} \hat{\sigma}_{gr_\mathrm{c}}^l ,\\
\label{eq:sigma_grc_dot}
&\dot{\hat{\sigma}}_{gr_\mathrm{c}}^l = -\Big(\gamma_r + i\delta  \Big) \hat{\sigma}_{gr_\mathrm{c}}^l + i\Omega_\mathrm{c}^* \hat{\sigma}_{ge_\mathrm{c}}^l ,
\end{align}

We again move to the frequency domain under the Fourier transformation (\ref{eq:four}).
Solving the control branch's set of EOMs and using the input-output relation for the control field $\hat{a}_\mathrm{in,c}+\hat{a}_\mathrm{out,c}=\sqrt{2\kappa_\mathrm{c}}\hat{a}_\mathrm{c}$, we obtain the reflection coefficient and the susceptibilities corresponding to transitions from the ground state to the excited state $\ket{e_\mathrm{c}}$ and to the Rydberg state $\ket{r_\mathrm{c}}$
\begin{align}
&R_\mathrm{c}[\omega]=\frac{\kappa_\mathrm{c}+i\omega-\frac{|g_\mathrm{c}|^2N}{\gamma_{e_\mathrm{c}}+i(\Delta-\omega)+\frac{|\Omega_\mathrm{c}|^2}{\gamma_r+i(\delta-\omega)}}}{\kappa_\mathrm{c}-i\omega+\frac{|g_\mathrm{c}|^2N}{\gamma_{e_\mathrm{c}}+i(\Delta-\omega)+\frac{|\Omega_\mathrm{c}|^2}{\gamma_r+i(\delta-\omega)}}},\label{eq:aprop}
\\&
\chi_{e_\mathrm{c}}[\omega] =\frac{\frac{ ig_\mathrm{c}}{(\gamma_{e_\mathrm{c}}+i(\Delta-\omega))+\frac{|\Omega_\mathrm{c}|^2}{(\gamma_r + i(\delta -\omega)}}}{\kappa_\mathrm{c}-i\omega+\frac{|g_\mathrm{c}|^2N}{\gamma_{e_{\mathrm{c}}}+i(\Delta-\omega)+\frac{|\Omega_{\mathrm{c}}|^2}{\gamma_r+i(\delta-\omega)}}},\label{eq:eprop}
\\&
\chi_{r_\mathrm{c}}[\omega] =\frac{\frac{ -\Omega^*_\mathrm{c} g_{\mathrm{c}}}{ (\gamma_{e_\mathrm{c}}+i(\Delta-\omega))(\gamma_r + i(\delta -\omega))+|\Omega_\mathrm{c}|^2}}{\kappa_\mathrm{c}-i\omega+\frac{|g_{\mathrm{c}}|^2N}{\gamma_{e_{\mathrm{c}}}+i(\Delta-\omega)+\frac{|\Omega_{\mathrm{c}}|^2}{\gamma_r+i(\delta-\omega)}}},
\label{eq:rprop}
\end{align}
respectively. These factors connect the control branch's operators with the control input field through $\hat{a}_\mathrm{out,c}=R_\mathrm{c}\hat{a}_\mathrm{in,c}$,  $\hat{\sigma}_{ge_\mathrm{c}}=\chi_{e_\mathrm{c}} \hat{a}_\mathrm{in,c}/\sqrt{2\kappa_\mathrm{c}}$,  $\hat{\sigma}_{gr_\mathrm{c}}=\chi_{r_\mathrm{c}}\hat{a}_\mathrm{in,c}/\sqrt{2\kappa_\mathrm{c}}$.

\par The conservation of probability dictates the balancing of the incoming and outgoing scattering processes of the system. For the control branch, this translates to the relation
\begin{align}
\notag
\langle (\hat{a}_\mathrm{in,c})^\dagger\hat{a}_\mathrm{in,c}\rangle=&\langle (\hat{a}_\mathrm{out,c})^\dagger\hat{a}_\mathrm{out,c}\rangle+\sum_l\langle(L_{ge_\mathrm{c}}^l)^\dagger L_{ge_\mathrm{c}}^l\rangle
\\&+\sum_l\langle(L_{r_\mathrm{c}}^l)^\dagger L_{r_\mathrm{c}}^l\rangle.\label{eq:1probcav}
\end{align}

By the use of the proportionality factors (\ref{eq:aprop})–(\ref{eq:rprop}), Eq. (\ref{eq:1probcav}) can be reexpressed as
\begin{equation}
|R_\mathrm{c}[\omega]|^2+\Gamma_{e_\mathrm{c}}[\omega]+\Gamma_{r_\mathrm{c}}[\omega]=1,
\end{equation}
which explicitly represents the conservation of probability. Here, we have introduced the loss probability through spontaneous emission of the control excited state $\ket{e_\mathrm{c}}$ given an incident photon $\Gamma_{e_\mathrm{c}}[\omega]=N\gamma_{e_\mathrm{c}}|\chi_{e_\mathrm{c}}[\omega]|^2/\kappa$, the dephasing probability of the Rydberg control state $\ket{r_\mathrm{c}}$ given an incident photon $\Gamma_{r_\mathrm{c}}[\omega]=\gamma_rN|\chi_{r_\mathrm{c}}[\omega]|^2/\kappa$  and the reflectance of the incoming control field $|R_\mathrm{c}[\omega]|^2$. 
\par Impedance matching is achieved when the dephasing probability of the Rydberg control state, derived by Eq. (\ref{eq:rprop}) as
\begin{align}
\Gamma_{r_\mathrm{c}}=\frac{4C_\mathrm{c}\frac{|\Omega_\mathrm{c}|^2\gamma_r}{\gamma_{e_\mathrm{c}}}}{\left|\frac{\gamma_r(1+C_\mathrm{c})\gamma_{e_\mathrm{c}}+\Delta\left(\frac{|\Omega_\mathrm{c}|^2}{\Delta}-\delta\right)}{\gamma_{e_\mathrm{c}}}+i\frac{\Delta\gamma_r+\delta(1+C_\mathrm{c})\gamma_{e_\mathrm{c}}}{\gamma_{e_\mathrm{c}}}\right|^2},
\end{align}
is close to unity, and accordingly, the reflectance and the decay rate through the excited state are close zero. In this case an incident control photon always induces a dephasing quantum jump, thereby storing the photon as a collective Rydberg excitation. To accomplish this the effective processes of the system need to be balanced and some of them suppressed. This is achieved by choosing the dephasing rate to be equal the effective output rate of the cavity
\begin{equation}
    \gamma_r=C_\mathrm{c}\gamma_{e_\mathrm{c}}|\Omega_\mathrm{c}|^2/\Delta^2,\label{eq:imc1}
\end{equation}
where $C_\mathrm{c}=|g_\mathrm{c}|^2N/(\kappa_\mathrm{c}\gamma_{e_\mathrm{c}})$ is the cooperativity of the control branch \cite{PhysRevLett.100.093603}.
Moreover, the detuning is chosen to be large compared with the effective decay rate from the control excited stated, i.e., $\Delta\gg(C_\mathrm{c}+1)\gamma_{e_\mathrm{c}}$. 
This condition is crucial because it ensures that the effective decay rate of the cavity is proportional to the number of atoms participating in the collective excitation \cite{Gorshkov2007a}. Consequently, localization will lead to suppression of this form of decay. A last condition is necessary in order to counter the ac Stark shift, that is present due to the control branch's  classical field $\Omega_\mathrm{c}$. This condition requires the two photon detuning to be equal to the ac Stark shift, i.e., 
$\delta=|\Omega_\mathrm{c}|^2/\Delta$. Under these conditions the dephasing probability of the Rydberg state given an incident photon on resonance is found to be
\begin{equation}
\Gamma_{r_\mathrm{c}}[\omega=0]=
\frac{1}{1+\frac{1}{C_\mathrm{c}}+\frac{1}{(2C_\mathrm{c})^2}},\label{eq:imc}
\end{equation}
which goes to unity for large values of the cooperativity $C_\mathrm{c}\gg1$, leading to fulfillment of the impedance matching requirement. Accordingly, the reflectance of the control field $R_\mathrm{c}$ and spontaneous emission loss probability from the excited state given an incident photon $\Gamma_{e_\mathrm{c}}$, go to zero in the same limit $C_\mathrm{c}\gg1$, since  
 \begin{align}
&|R_\mathrm{c}[\omega=0]|^2=\frac{1}{(2C_\mathrm{c}+1)^2},
\\&
\Gamma_{e_\mathrm{c}}[\omega=0]=\frac{4C_\mathrm{c}}{(2C_\mathrm{c}+1)^2}.
\end{align}
Under these conditions the control photon is transferred into a collective Rydberg excitation. This analytical estimate is confirmed numerically in Sec. \ref{sec:rescav}.


\subsection{\label{sec:simucav}Numerical simulation}
In this section the wave-function Monte Carlo (wfMC) approach \cite{Dalibard1992,Molmer1993} is introduced to simulate the system of a  Rydberg atomic ensemble located in a single-sided cavity. We considered $N$=1000 atoms located inside the cavity and three different atomic distributions were examined. The first distribution was a symmetric ring geometry,  where the neighboring atoms were equidistant. 
The constant neighboring distance leads to a constant blockaded cooperativity $C_\mathrm{b,p}^k$ for $k\in[1,N]$, which consequently leads to identical dephasing rates
$\gamma^k_r$, $k\in[1,N]$ for all atoms. The second and third distributions investigated were more realistic versions, where the atoms were randomly distributed in space following a Gaussian distribution in one and three dimensions. The two-dimensional version was not included as it is not relevant to conventional experimental setups. In these more realistic models the blockaded cooperativity depends on the position of each atom, and consequently so does the dephasing rate. Comparing the results between the realistic and symmetric models reveals the impact of distance fluctuations on device performance. For all the geometries we assume that the atoms have identical coupling constants to each field.
\par The full system is described by the control Hamiltonian (\ref{eq:H_control}), two decay operators (\ref{eq:kjum}) and (\ref{eq:gjum}) and two dephasing operators (\ref{eq:kjump}) and (\ref{eq:gjump}).
The two dephasing operators effectively describe the probe branch, which has been adiabatically eliminated as discussed in Sec. \ref{sec:probcav}. 
The corresponding non-Hermitian Hamiltonian of the effectively described system reads
\begin{align}\notag
\hat{\mathcal{H}}_\mathrm{NH}=&\hat{\mathcal{H}}_\mathrm{\mathrm{control}}-\frac{i}{2}\hat{L}_{\kappa_\mathrm{c}}^\dagger\hat{L}_{\kappa_\mathrm{c}}-\frac{i}{2}\sum_{l=1}^N(\hat{L}^l_{ge_\mathrm{c}})^\dagger\hat{L}^l_{ge_\mathrm{c}}
\\&-\frac{i}{2}\hat{L}_{\kappa_\mathrm{p}}^\dagger\hat{L}_{\kappa_\mathrm{p}}-\frac{i}{2}\sum_{l=1}^N(\hat{L}_{ge_\mathrm{p}}^l)^\dagger\hat{L}_{ge_\mathrm{p}}^l
\end{align}
The basis of the Hilbert space associated with the system is given by the $(2N+1)$-dimensional vector  $\{\ket{g,1},\ket{e_\mathrm{c}^1},...,\ket{e_\mathrm{c}^N},\ket{r_\mathrm{c}^1},...,\ket{r_\mathrm{c}^N}\}$ because we allow the presence of only a single control photon. The non-Hermitian Hamiltonian projected on this basis can be written in matrix form as the $(2N+1)
\times (2N+1)$ matrix
\newcommand\scalemath[2]{\scalebox{#1}{\mbox{\ensuremath{\displaystyle #2}}}}

\[
\bold{H_{NH}}= \left(
    \scalemath{0.63}{
\begin{array}{ccccccccc}
 -i \kappa_\mathrm{c}  & g_\mathrm{c} & g_\mathrm{c} & \cdots & g_\mathrm{c} & 0 & \cdots & \cdots& 0  \\
 g_\mathrm{c} & \Delta -i \gamma_{e_\mathrm{c}}  & 0 & \cdots & 0  & \Omega_\mathrm{c} & \ddots & & \vdots   \\
 g_\mathrm{c} & 0  & \Delta -i \gamma_{e_\mathrm{c}} & 0  & \cdots &0 & \Omega_\mathrm{c} & \ddots & \vdots  \\
 \vdots & \vdots & \ddots & \ddots &\ddots &\vdots & \ddots & \ddots & 0 \\
 g_\mathrm{c} & 0 & \cdots  &0& \Delta -i \gamma_{e_\mathrm{c}} & 0  & \cdots & 0& \Omega_\mathrm{c} \\
 0 & \Omega_\mathrm{c} & 0  & \cdots & 0 & \delta - i\gamma_r^1 & 0 & \cdots& 0 \\
 \vdots & \ddots & \Omega_\mathrm{c}  & 0 & \cdots& 0&\delta - i\gamma_r^2 & \ddots & \vdots \\
 \vdots &    &  \ddots  & \ddots & \ddots &\vdots & \ddots & \ddots & 0\\
 0 & \cdots & \cdots   & 0 & \Omega_\mathrm{c} & 0& \cdots& 0 & \delta - i\gamma_r^N \\
\end{array}
    }
  \right)
\]
where $\gamma_r^k= |\gamma_{\kappa_\mathrm{c}}^k|+\sum_l^N|\gamma_{ge_\mathrm{c}}^{k,l}|$ is the effective dephasing rate resulting from decay processes on the probe branch (\ref{eq:kjump}) and (\ref{eq:gjump}).

\par In a similar manner, the wave-function describing the system at time $t$ is expanded as
\begin{equation}
\ket{\Psi(t)}=c_g\ket{g,1}+\sum_{l=1}^N c^l_{e_\mathrm{c}}(t)\ket{e_\mathrm{c}}_l+\sum_{l=1}^N c^l_{r_\mathrm{c}}(t)\ket{r_\mathrm{c}}_l.
\end{equation}
Initially, the system is deexcited meaning that $c_g(t_0)=c^l_{e_\mathrm{c}}(t_0)=c^l_{r_\mathrm{c}}(t_0)=0$, $l\in[0,N]$. The control excitation is introduced to the system by a long single-photon pulse of Gaussian profile that starts entering the system at $t=t_0$. This is described by
\begin{equation}
c_\mathrm{in}(t)=\frac{1}{4\pi\sigma}e^{-(t-t_0-t_\mathrm{m})^2/(4\sigma^2)}.\label{eq:cin}
\end{equation}
The input control-photon pulse is normalized according to $
\int^{t_\mathrm{max}}_{t_\mathrm{0}}dt|c_\mathrm{in}(t)|^2=1$. The values used for the simulation are $\sigma=160/\gamma_{e_\mathrm{p}}$, $t_\mathrm{m} =500/\gamma_{e_\mathrm{p}}$, $t_0 =0$ and the total duration of the pulse is $t_\mathrm{tot} =1000/\gamma_{e_\mathrm{p}}$.
\par The system evolves under the non-Hermitian Hamiltonian and in the presence of the control-photon pulse the evolution is given by
\begin{equation}\label{eq:sim}
\frac{d}{dt}\ket{\Psi(t)}=-i \bold{H_{NH}} \ket{\Psi(t)}+\sqrt{2\kappa_\mathrm{c}}c_\mathrm{in}(t)\ket{g,1}.
\end{equation}
The total norm of the system's wave-function and the control-photon pulse at time $t$ is
\begin{align}\notag
&\langle\Psi(t)|\Psi(t)\rangle+\int_t^{t_\mathrm{max}}|c_\mathrm{in}(t)|^2=|c_g(t)|^2\\&+\sum^N_l|c_{e_\mathrm{c}}(t)|^2+\sum^N_l|c_{r_\mathrm{c}}(t)|^2+\int_t^{t_\mathrm{max}}|c_\mathrm{in}(t)|^2.
\end{align}
The norm's initial value is unity at $t=t_0$ and is reduced under the evolution of the non-Hermitian dynamics, while the control-photon pulse is fed into the dissipative system.
\par As the first step of the process a quantum jump will occur since an incident photon will either be subject to a dephasing quantum jump or the photon will leave the system corresponding to a decay jump. The time of the jump is decided by the standard stochastic  wave-function Monte Carlo (wfMC) procedure: A value between 0 and 1 is randomly chosen and the time of the jump set to the temporal point, when the norm reaches that value.

\par Once the time of jump $t_j$ is set, a second stochastic process determines the nature of the jump as one of the four jump operators (\ref{eq:kjum}),(\ref{eq:gjum}),(\ref{eq:kjump}) or (\ref{eq:gjump}). The non-normalized probabilities of these jumps are calculated as
\begin{align}
&p_{\gamma_{ge_\mathrm{c}}}(t_{j})=2\gamma_{e_\mathrm{c}}\sum_{k=1}^N\left|c_{e_\mathrm{c}}^k(t_{j})\right|^2, \label{eq:pegc}
\\&p_{\kappa_\mathrm{c}}(t_{j})=\left|\sqrt{2\kappa_\mathrm{c}}c_g(t_{j})-c_\mathrm{in}(t_{j})\right|^2, \label{eq:pkc}
\\&p_{\kappa_\mathrm{p}}(t_{j})=\sum_{k=1}^N|\gamma_{\kappa_\mathrm{p}}^k||c^k_{r_\mathrm{c}}(t_{j})|^2, \label{eq:pkp}
\\&p_{\gamma_{ge_\mathrm{p}}}(t_{j})
=\sum_{l=1}^N\sum_{\substack{k=1\\k\neq l}}^N|\gamma_{e_\mathrm{p}}^{k,l}||c^k_{r_\mathrm{c}}(t_{j})|^2,\label{eq:pegp}
\end{align}
where the dephasing rates $\gamma_{\kappa_\mathrm{p}}^k$, $\gamma_{e_\mathrm{p}}^{k,l}$ were introduced in Eqs. (\ref{eq:kjump}) and (\ref{eq:gjump}).
Additionally, the normalized probabilities $\Pi_i$ of the jumps are found via a normalization process $\Pi_i=p_i(t_j)/\sum_ip_i(t_j)$.
Depending on the nature of the jump the evolution of the system takes different paths:

\begin{figure}[t!]
\includegraphics[width=1.\columnwidth]{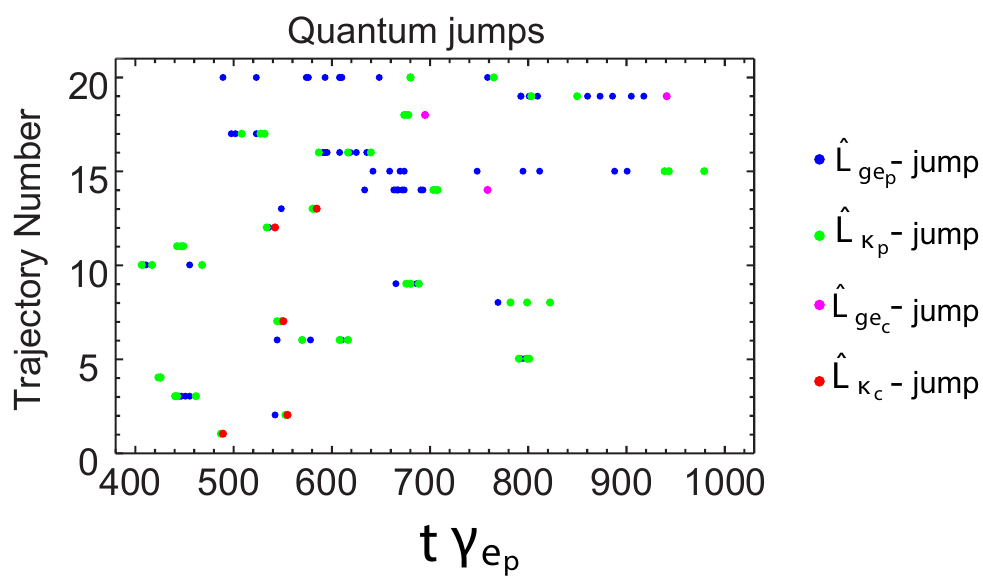}
\caption{First 20 trajectories of the Monte Carlo simulation for 3D Gaussian atomic distribution of $N=1000$ atoms in a cavity with cooperativity $C_\mathrm{c}=100$ and blockaded cooperativity $\overline{C}_\mathrm{b,p}=0.5$. The color of the dot indicate the nature of the jump that occur at the specific time. The parameters are fixed to $\Delta/\gamma_{e_\mathrm{c}} = 180, \kappa_\mathrm{p}=\kappa_\mathrm{c}=\gamma_{e_\mathrm{p}}=\gamma_{e_\mathrm{c}}, \Omega_\mathrm{c}/\gamma_{e_\mathrm{c}} = 5, \Omega_\mathrm{p}/\gamma_{e_\mathrm{c}}=10$, $\delta/\gamma_{e_\mathrm{c}}=1.09$, $|\alpha_{\mathrm{in,p}}|^2/\gamma_{e_\mathrm{c}}=0.33$.}
\label{fig:cavjum}
\end{figure}

\par\textit{Decay jumps $\mathit{\hat{L}_{\kappa_\mathrm{c}}}$, $\mathit{\hat{L}_{\gamma_{ge_\mathrm{c}}}}$.---} In the cases of one of these two decay jumps, the control photon is lost into the environment by spontaneous emission of one atom from the excited state $\ket{e_\mathrm{c}}$ or via decay of the cavity. This type of decay jump concludes the trajectory and the simulation comes to an end.
\par\textit{Dephasing jump $\mathit{\hat{L}_{\gamma_{\kappa_\mathrm{p}}}}$.---} In the occurrence of a dephasing jump due a change in cavity reflection of the probe, the control input photon is fully absorbed by the atoms and the state of the system becomes
\begin{equation}
\ket{\Psi'(t_{j})}=\frac{\hat{L}_{\kappa_\mathrm{p}}\ket{\Psi(t_{j})}}{\sqrt{p_{\kappa_\mathrm{p}}(t_{j})}}.
\end{equation}
Then the process is repeated by solving Eq. (\ref{eq:sim}) without the single-photon pulse term, with the subsequent jump determined by the same procedure.
\par\textit{Dephasing jump $\mathit{\hat{L}_{\gamma_{ge_\mathrm{p}}}}$.---} In the case of a dephasing jump due to the spontaneous emission of an excited atom in the probe branch, a third stochastic process is necessary in order to identify which of the $N$ atoms decayed from the probe branch's excited state $\ket{e_\mathrm{p}}$. The normalized probability of the $l$th atom to decay is
\begin{equation}
\Pi_{\gamma_{ge_\mathrm{p}}}^l(t_{j})=\frac{p^l_{\gamma_{ge_\mathrm{p}}}(t_{j})}{\sum_{l=1}^Np^l_{\gamma_{ge_\mathrm{p}}}(t_{j})}, \label{eq:pi}
\end{equation}
where $p^l_{\gamma_{ge_\mathrm{p}}}(t_{j})=\sum_{\substack{k=1\\k\neq l}}^N|\gamma_{ge_\mathrm{p}}^{k,l}||c^k_{r_\mathrm{c}}(t_{j})|^2$.
Once the atom that decayed has been determined, the control-photon pulse is fully absorbed by the system and the system is prepared in the state
\begin{equation}
\ket{\Psi'(t_{j})}=\frac{\hat{L}^l_{\gamma_{ge_\mathrm{p}}}\ket{\Psi(t_{j})}}{\sqrt{p^l_{\gamma_{ge_\mathrm{p}}}(t_{j})}},\label{eq:psi}
\end{equation}
which is localized around the $l$th atom that decayed, as explained in the Sec. \ref{sec:probcav}. Then the process is repeated for determining the time and the nature of the next jump.
\par The process is repeated until the control excitation decays through one of the two decay jumps ($\hat{L}_{ge_\mathrm{c}}$, $\hat{L}_{\kappa_\mathrm{c}}$) or we determine that sufficiently many $\hat{L}_{\kappa_\mathrm{p}}$ dephasing jumps have occurred that it constitutes a successful switching event.  
\par Our results are averaged over $N_\mathrm{traj}=2000$ trajectories for each value of the cooperativity $C_\mathrm{c}$ of the control branch and the average blockaded cooperativity $\overline{C}_\mathrm{b,p}=1/N\sum_k^NC_\mathrm{b,p}^k$ of the probe branch. In Fig. \ref{fig:cavjum} the first 20 trajectories of a simulation with $ C_\mathrm{c}=100$ and $\overline{C}_\mathrm{b,p}=0.5$ for a three-dimensional (3D) atomic Gaussian distribution are depicted. Each trajectory comes to an end, with either successful $\hat{L}_{\kappa_\mathrm{p}}$ dephasing jumps (green dots) or is unsuccessful with one of the decay jumps (pink and red dots). Furthermore it can be observed that the occurrence of an $\hat{L}_{ge_\mathrm{p}}$ dephasing jump is rarely followed by an $\hat{L}_{\kappa_\mathrm{c}}$ decay jump, since the localization induced by the dephasing jump strongly suppresses this type of decay as explained below.


\subsection{\label{sec:rescav}Results}

\begin{figure}[t!]
\includegraphics[width=1.03\columnwidth]{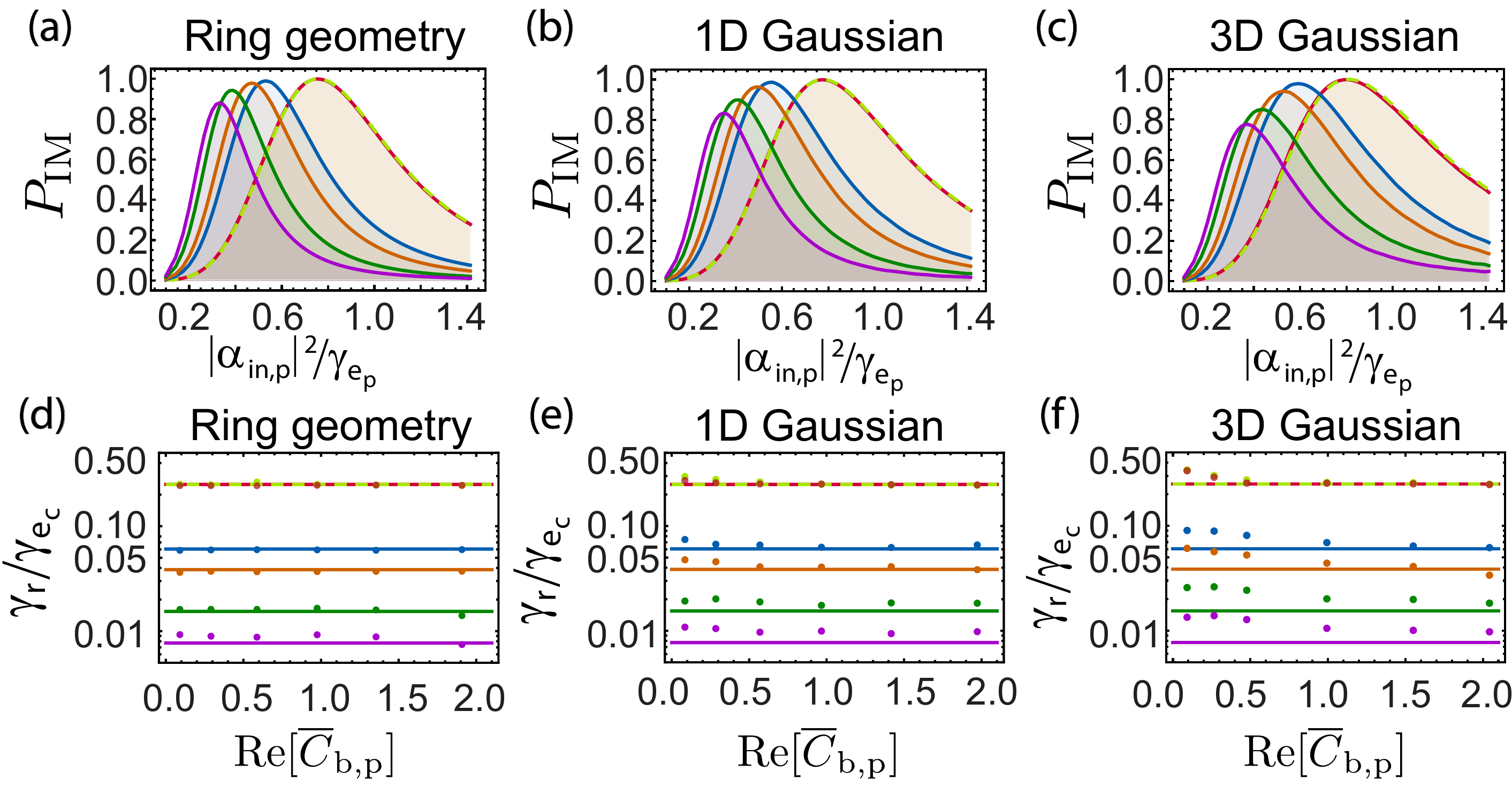}
\caption{Impedance matching results for the cavity model with $N=1000$ atoms. (a)-(c) Impedance matching probability $P_{\mathrm{IM}}$ as a function of probe strength for $\mathrm{Re}[\overline{C}_\mathrm{b,p}]\approx0.5$, with multiple curves corresponding to different values of cooperativity $C_\mathrm{c}= 10, 20, 50, 100, 1000, 5000$, shown for different atomic distributions: (a) symmetric ring geometry (b) 1D Gaussian (c) 3D Gaussian. (d)-(f) Numerically optimized dephasing rate (dots) and the theoretical estimate (solid line) versus the blockaded cooperativity, shown for the same values of $C_\mathrm{c}$ and atomic distributions: (d) symmetric ring geometry (e) 1D Gaussian (f) 3D Gaussian. In all panels, lower curves correspond to lower $C_\mathrm{c}$, and the curves for $C_\mathrm{c} = 1000$ and $5000$ nearly overlap. The parameters are fixed to $ \kappa_\mathrm{p}=\kappa_\mathrm{c}=\gamma_{e_\mathrm{p}}=\gamma_{e_\mathrm{c}}, \Omega_\mathrm{p}/\gamma_{e_\mathrm{c}}=10, \Delta/\gamma_{e_\mathrm{c}} = 180,\Omega_\mathrm{c}/\gamma_{e_\mathrm{c}} = 5$ for $C_\mathrm{c}=10,20,50,100$, and  $\Delta/\gamma_{e_\mathrm{c}} = 4C_\mathrm{c}/5, \Omega_\mathrm{c}/\gamma_{e_\mathrm{c}} =20,45,$ for $ C_\mathrm{c}=1000,5000$. The detuning $\delta$ is optimized at each point to enhance $P_{\mathrm{IM}}$. } 
\label{fig:im}
\end{figure}

The results section is organized in three parts. The first focuses on the impedance-matching conditions, which are numerically optimized and compared with our analytical estimate derived in the Sec. \ref{subsec:imcav}. The second part presents the numerical optimization of the efficiency of the SPT versus the  cooperativity $C_\mathrm{c}$ of the control branch and the average blockaded cooperativity $\overline{C}_\mathrm{b,p}$ of the probe branch. Finally, in the third part we discuss the effects that limit the efficiency and possible improvements.
\par\textit{Impedance matching.---}The impedance matching probability $P_{\mathrm{IM}}$ is defined as the ratio of the probability of the first jump being a dephasing jump ($\hat{L}_{{ge_\mathrm{p}}}$, $\hat{L}_{\kappa_\mathrm{p}}$), instead of a decay jump ($\hat{L}_{ge_\mathrm{c}}$, $\hat{L}_{\kappa_\mathrm{c}}$). This is given by the relation
\begin{equation}
P_{\mathrm{IM}}=\frac{\int_{t_0}^{t_{max}}dt\left(p_{\gamma_{\kappa_\mathrm{p}}}(t)+p_{\gamma_{ge_\mathrm{p}}}(t)\right)}{\int_{t_0}^{t_{max}}dt\left(p_{\gamma_{ge_\mathrm{c}}}(t)+p_{\kappa_\mathrm{c}}(t)+p_{\gamma_{\kappa_\mathrm{p}}}(t)+p_{\gamma_{ge_\mathrm{p}}}(t)\right)},
\end{equation}
where the probabilities for the realization of each jump are defined in Eqs. (\ref{eq:pegc})–(\ref{eq:pegp}). It is essential for the function of the SPT that $P_{\mathrm{IM}}$ is close to unity because it sets the upper limit for the efficiency, since it guarantees that the control photon is absorbed by the atomic ensemble.
\par As shown in Fig. \ref{fig:im}(a)–\ref{fig:im}(c) the optimization is done by varying the strength of the input probe field for different atomic distributions. The parameters can be found in the caption. The simulations are repeated for different values of the control branch's cooperativity $C_\mathrm{c}=10,20,50,100,500,1000$ and blockaded cooperativity $\mathrm{Re}[\overline{C}_\mathrm{b,p}]\approx0.1,0.25,0.5,1,1.5,2$ for the symmetric ring geometry, the one-dimensional (1D) Gaussian and the 3D Gaussian atomic distribution.

\begin{figure}[t!]
\includegraphics[width=1.\columnwidth]{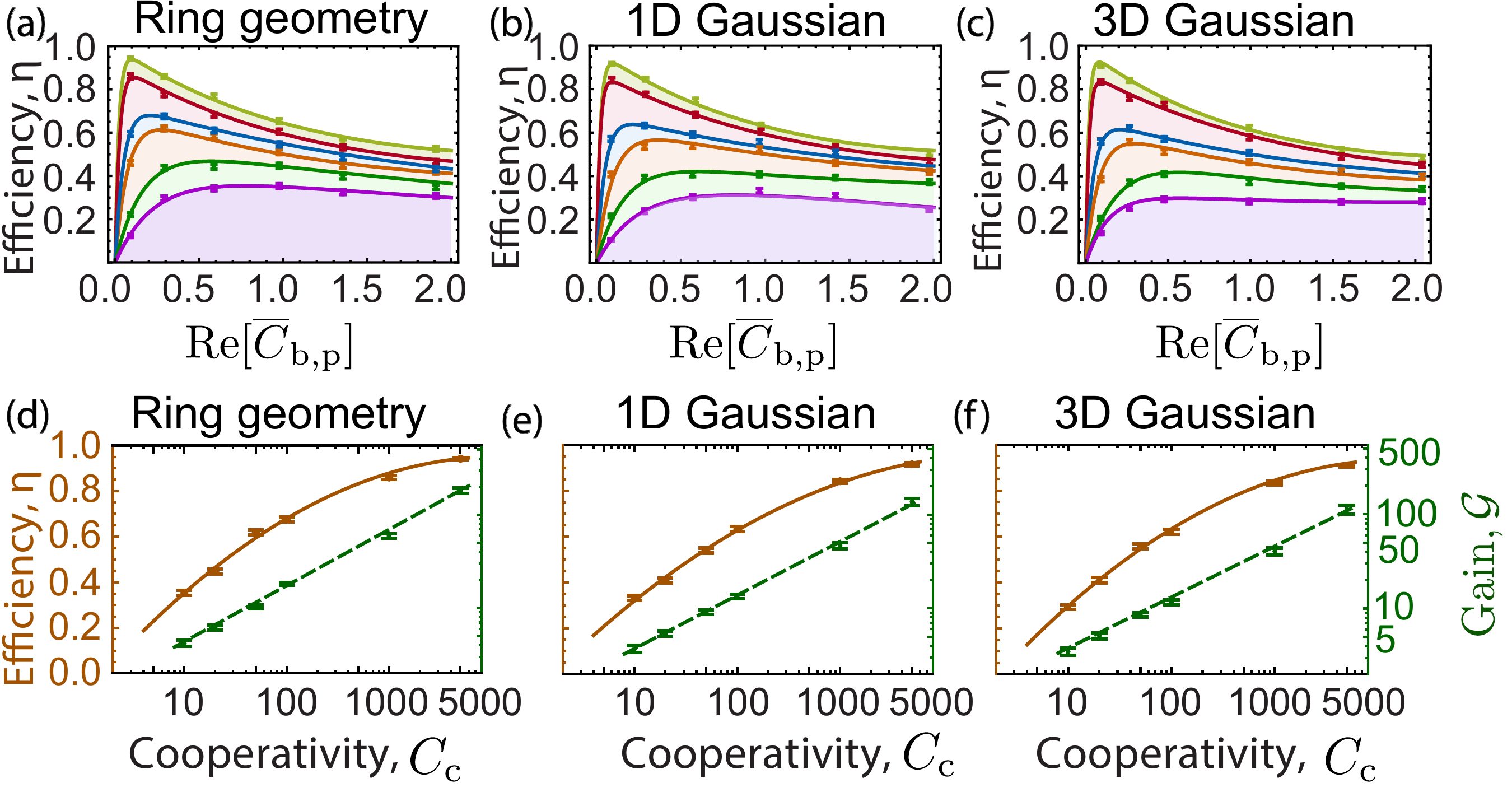}
\caption{Performance of the cavity model with $N = 1000$ atoms.
(a)–(c) SPT efficiency as a function of blockaded cooperativity, for control branch cooperativity values $C_\mathrm{c} = 10, 20, 50, 100, 1000, 5000$ (lower curves correspond to lower $C_\mathrm{c}$). Results are shown for different atomic distributions:
(a) symmetric ring geometry, (b) 1D Gaussian, (c) 3D Gaussian.
(d)–(f) SPT efficiency (solid) and gain (dashed) as functions of cooperativity, with the blockaded cooperativity optimized at each point. Results are shown for different atomic distributions:
(d) symmetric ring geometry, (e) 1D Gaussian, (f) 3D Gaussian.
The parameters are fixed to $\kappa_\mathrm{p} = \kappa_\mathrm{c} = \gamma_{e_\mathrm{p}} = \gamma_{e_\mathrm{c}}$, with $\Omega_\mathrm{p}/\gamma_{e_\mathrm{c}} = 10$, $\Delta/\gamma_{e_\mathrm{c}} = 180$, and $\Omega_\mathrm{c}/\gamma_{e_\mathrm{c}} = 5$ for $C_\mathrm{c} = 10, 20, 50, 100$; and $\Delta/\gamma_{e_\mathrm{c}} = 4C_\mathrm{c}/5$, $\Omega_\mathrm{c}/\gamma_{e_\mathrm{c}} = 20, 45$ for $C_\mathrm{c} = 1000, 5000$. The detuning $\delta$ and probe strength $|\alpha_{\mathrm{in,p}}|^2$ are optimized at each point to enhance efficiency. Solid curves are polynomial fits provided as guides to the eye; points correspond to simulation data.}
\label{fig:jk}
\end{figure}

\par The value of $P_\mathrm{IM}$ is observed to be in excellent agreement with the analytical estimate given in Eq.~(\ref{eq:imc}). The maximum value of $P_\mathrm{IM}$ is obtained for $C_\mathrm{c} = 5000$, with $P_\mathrm{IM} = 0.9997$ for the symmetric ring geometry, $0.9996$ for the 1D Gaussian distribution, and $0.9994$ for the 3D Gaussian distribution. The result for the symmetric ring geometry is closest to the theoretical prediction of $0.9998$, while the slight deviations observed in the 1D and 3D Gaussian distributions are due to the reduced symmetry present in these more realistic atomic configurations.
\par Furthermore, in Figs. \ref{fig:im}(d)–\ref{fig:im}(f) the numerically optimized impedance matching conditions for the dephasing rate $\gamma_r^\mathrm{opt}=\sum_k^N(\gamma_{\kappa_\mathrm{p}}^k+\sum_l^N\gamma_{e_\mathrm{p}}^{k,l})$ (dots) are compared with the theoretical estimate (solid line) in Eq. (\ref{eq:imc1}). For the symmetric ring case we observe a very close agreement between the two, while as the system becomes more realistic, we observe a small deviation due to the inhomogeneity of the system. For lower values of $\overline{C}_
\mathrm{b,p}$ the observed deviation is due to a broad plateau of almost equal $P_\mathrm{IM}$ probabilities, where the highest value was chosen.

\par\textit{Efficiency.---}The final readout of the device can be obtained by counting photons or with a homo- or heterodyne detection scheme, which measures the difference in the output field due to the Rydberg blockade mechanism of the control photon. Alternatively the SPT can be incorporated into an interferometric setup, which  cancels the  probe signal on a photodetector in the absence of control photons (note that in this case the output is in a coherent state and can hence be canceled exactly). Since the reflection operator $\hat{L}_{\kappa_\mathrm{p}}$ shown in Eq. \eqref{eq:kjump} exactly corresponds to the difference  between having a control photon or not, a detection on such a detector exactly correspond to a $\hat{L}_{\kappa_\mathrm{p}}$ jump. For simplicity we thus characterize the performance of the proposed SPT device by the number of $\hat{L}_{\kappa_\mathrm{p}}$ jumps, noting that similar information can also be extracted using homo- or heterodyne detection. In practice we require the number of detection jumps $\hat{L}_{\kappa_\mathrm{p}}$ to be larger than or equal to a fixed threshold number which we take to be   $N_{\gamma_{\kappa_\mathrm{p}}}^{th} = 3$. We then define the efficiency as the probability for this to happen as the corresponding fraction of trajectories
\begin{equation}
\eta=\frac{N_\mathrm{traj}(N_{\gamma_{\kappa_\mathrm{p}}}^{th}\geq3)}{N_\mathrm{traj}}.\label{eq:effc}
\end{equation}
For a homodyne measurement scheme this requirement roughly corresponds to a homodyne signal whose squared amplitude is six times the vacuum noise if we disregard complications due to the multimode nature of the outgoing field.

\textit{Gain.---}The gain of the transistor is defined as the average number of probe photons scattered due to a single stored control excitation, resulting in a measurable change in probe reflection through the cavity. It is obtained by counting the number of $\hat{L}_{\kappa_\mathrm{p}}$ jumps per trajectory, each corresponding to a change in reflection due to disrupted EIT. Since the number of such jumps fluctuates between trajectories, the total gain $\mathcal{G}$ is defined as the average over all trajectories contributing to the efficiency.
\par The simulations are repeated for different values of the control branch's cooperativity $C_\mathrm{c}=10,20,50,100,1000,5000$ and blockaded cooperativity $\mathrm{Re}[\overline{C}_\mathrm{b,p}]\approx0.1,0.25,0.5,1,1.5,2$ for the symmetric ring geometry, the 1D Gaussian and the 3D Gaussian atomic distribution. The performance results are shown in Fig.~\ref{fig:jk}. The maximum efficiency is achieved for the highest considered value of the cooperativity $C_\mathrm{c} = 5000$ and $\mathrm{Re}[\overline{C}_\mathrm{b,p}] = 0.09$, reaching $\eta = 0.942$ with an associated gain of $\mathcal{G} = 181$ for the symmetric ring geometry. For the 1D Gaussian distribution, the maximum efficiency is $\eta = 0.916$ with $\mathcal{G} = 136$, and for the 3D Gaussian distribution, $\eta = 0.908$ with $\mathcal{G} = 113$. The reduction in efficiency and gain observed in Gaussian distributions compared with the symmetric case is due to Rydberg interaction-induced dephasing caused by distance fluctuations.
\par Higher values of the control branch's cooperativity are observed to result in increased efficiency. Although similar experiments typically demonstrate cooperativity values on the order of a few tens~\cite{Vaneecloo2022,PhysRevX.12.021035}, the values used in our simulations have been achieved experimentally~\cite{Sauer2004}. We note that the upper limit of the observed efficiency is constrained by the capabilities of numerical simulations rather than by physical feasibility. This indicates that, although current experimental setups achieve only limited cooperativity, the underlying physical principles could support much higher efficiencies with greater cooperativity, which future devices are anticipated to attain. Relevant physical values and experimental parameters are discussed in Appendix \ref{appendix:b}.

\textit{Limitation of efficiency.---}The efficiency of the device is limited by the deleterious jumps responsible for the unsuccessful trajectories, which are the decay $\hat{L}_{ge_\mathrm{c}}$ or $\hat{L}_{\kappa_\mathrm{c}}$ jumps related to the loss of the control-photon excitation due to the decay of an excited atom and loss through the cavity, respectively. For the considered operating conditions the decay through the cavity $\hat{L}_{\kappa_\mathrm{c}}$ jumps happen at a much faster rate, thus limiting the lifetime of the stored excitation.

\begin{figure}[t!]
\includegraphics[width=1.02\columnwidth]{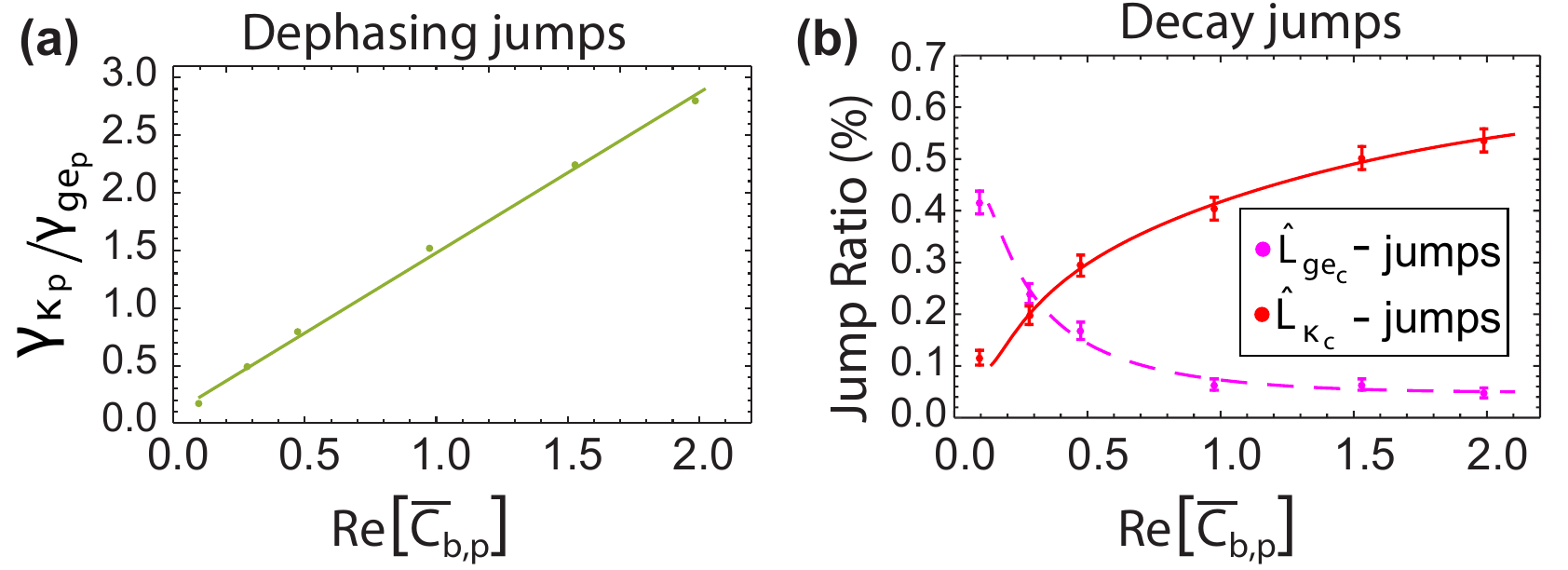}
\caption{Dissipative processes for the cavity model of $N = 1000$ atoms with a 3D Gaussian distribution.  (a) Ratio of the two dephasing rates versus the blockaded cooperativity for a cooperativity $C_\mathrm{c} = 100$. (b) Percentage of unsuccessful trajectories due to spontaneous emission from the probe excited state (dashed line) and decay of the cavity the cavity (solid line) versus the blockaded cooperativity for cooperativity $C_\mathrm{c} = 100$. Solid and dashed curves show fitted polynomials
and serve as a guide to the eye. The parameters are fixed to $\kappa_\mathrm{p} = \kappa_\mathrm{c} = \gamma_{e_\mathrm{p}} = \gamma_{e_\mathrm{c}}$, with $\Omega_\mathrm{p}/\gamma_{e_\mathrm{c}} = 10$, $\Delta/\gamma_{e_\mathrm{c}} = 180$, $\Omega_\mathrm{c}/\gamma_{e_\mathrm{c}} = 5$, and $\delta$ varies between $0.14\gamma_{e_\mathrm{c}}$ and $0.1\gamma_{e_\mathrm{c}}$ to provide better impedance matching and $|\alpha_\mathrm{in,p}|^2$ was chosen to optimize impedance matching at each point.}
\label{fig:deph}
\end{figure}

On one hand, the localization plays an essential role in the functioning of the SPT because it enhances the lifetime of the stored excitation by suppressing the $\hat{L}_{\kappa_\mathrm{c}}$ jumps. This is achieved through $\hat{L}_{ge_\mathrm{p}}^l$ jumps by the following mechanism: The storage dynamics transfer the excitation to a collective superposition of  all states in $\ket{e_\mathrm{c}}$, which has an enhanced decay rate $C_\mathrm{c}\gamma_{e_\mathrm{c}}$. This leads to the input-output rate $\gamma_{\mathrm{out}}$ being enhanced by a factor of $C_{\mathrm{c}}\propto N$ \cite{PhysRevLett.100.093603}.  When the excitation localizes due to a spontaneous emission jump in the probe branch $\hat{L}_{ge_\mathrm{p}}^l$, revealing information about the location of the excitation on the control branch $\ket{r^k_\mathrm{c}}$, the number of atoms participating in the superposition goes down, leading to a reduced decay rate and a longer lifetime. For high values of $\overline{C}_\mathrm{b,p}$, the cavity is fully blocked and we only have $\hat{L}_{\kappa_\mathrm{c}}$ jumps, as shown in Fig. \ref{fig:deph}(a). This means that the control excitation is not localised and decay rapidly through $\hat{L}_{\kappa_\mathrm{c}}$ jumps, as shown in Fig. \ref{fig:deph}(b).  

On the other hand, for low $\overline{C}_\mathrm{b,p}$ we do have very strong localization and the excitation lives long until it is finally lost by spontaneous emission through an $\hat{L}_{ge_\mathrm{c}}$ jump [Fig. \ref{fig:deph}(b)]. However, a problem arises when $\overline{C}_\mathrm{b,p}$ is small, despite the prolonged lifetime of the excitation, its impact on the probe's reflection is negligible, making it practically invisible to detection. Since for small $\overline{C}_\mathrm{b,p}$ the dephasing $\hat{L}_{\kappa_\mathrm{p}}$ jumps are suppressed, as shown in Fig. \ref{fig:deph}(a).

As a consequence of the above an optimization process is necessary to balance the relevant processes and the decay rates. The optimal $\overline{C}_\mathrm{b,p}$ is found at the point where the two sources of decay jumps are equal, as observed when comparing Fig.~\ref{fig:jk}(c) with Fig.~\ref{fig:deph}(b).
\par A higher value of the cooperativity $C_\mathrm{c}$ is beneficial for the functioning of the device, since the localization does not lead to such a dramatic decrease of $\overline{C}_\mathrm{b,p}$. This results from the fact that the same number of localized atoms will have a higher $\overline{C}_\mathrm{b,p}$, thus the dephasing $\hat{L}_{\kappa_\mathrm{p}}$ jumps occur more frequently, while the excitation retains the benefit of an enhanced lifetime due to localization. As a consequence improvement of the efficiency can be achieved by the implementation of a cavity with higher cooperativity $C_\mathrm{c}$, although this may be experimentally challenging. This limitation is not present for the free space model, analyzed in Sec. \ref{sec:fs}. 


\section{\label{sec:fs}Free Space}

\subsection{\label{sec:sysfs}System}

\begin{figure}[t!]
\includegraphics[width=1.\columnwidth]{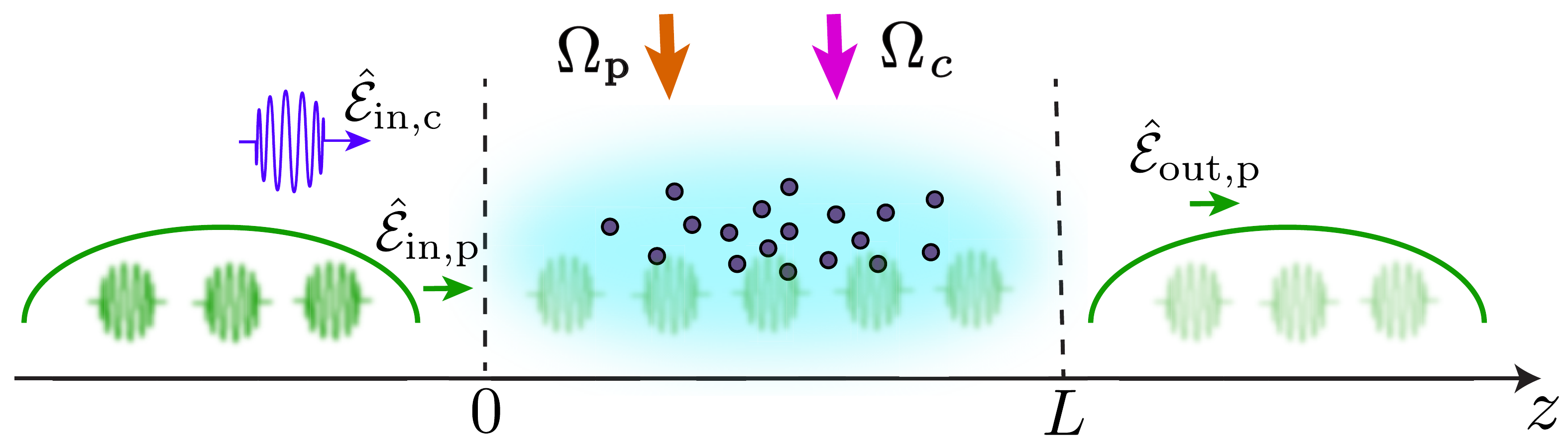}
\caption{Sketch of the free space device configuration, where the probe propagates along the $z$ axis inside the Rydberg cloud, which fills  part of the system $\{ 0,L\}$.}
\label{fig:fssketch}
\end{figure}

The second system we consider also consists of a Rydberg atomic cloud, but this time located in free space. As in the cavity case the atomic cloud is subject to two driving fields and two weak fields, one serving as a probe and the other as a control field. The system is sketched in Fig.~\ref{fig:fssketch}, and assumes propagation of both probe and control signals along the $z$-axis in the Rydberg medium. $N$ atoms are placed in 1D over a length $L$. The relevant energy structure of the Rydberg atoms is the same as for the cavity case and can be seen in Fig.~\ref{fig:sketch}(b).  The corresponding Hamiltonian consists of several parts, similar to the cavity case,
\begin{align}
\label{eq:Hfs}
&\hat{\mathcal{H}} = \hat{\mathcal{H}}_{\mathrm{probe}} + \hat{\mathcal{H}}_{\mathrm{control}} + \hat{\mathcal{H}}_{\mathrm{Ryd}} + \hat{\mathcal{H}}_{\mathrm{propag}} + \hat{\mathcal{H}}_{\mathrm{res}},
\end{align}
where instead of $\hat{\mathcal{H}}_{\mathrm{input}}$ describing the input and output for the cavity, we have the propagation of the field through the medium $\hat{\mathcal{H}}_{\mathrm{propag}}$. The probe part reads
\begin{align}
\label{eq:H_probefs}\notag
\hat{\mathcal{H}}_{\mathrm{probe}}=&-\hbar\int_{0}^{L}dzn(z) (g_\mathrm{p} \hat{\mathcal{E}}_\mathrm{p}(z,t) \hat{\sigma}_{e_\mathrm{p}g}(z,t)\\&+\Omega_\mathrm{p}\hat{\sigma}_{r_\mathrm{p}e_\mathrm{p}}(z,t)+ \mathrm{H.c.}) ,
\end{align}
where a traveling electromagnetic (EM) probe field at location $z$ is described by the creation (annihilation) operator $\hat{\mathcal{E}}^{\dagger}_{\mathrm{p}}$ ($\hat{\mathcal{E}}_\mathrm{p}$), and couples to the atoms with coupling constant $g_{\mathrm{p}}$ and the classical drive couples the $|e_\mathrm{p}|\rangle \leftrightarrow |r_\mathrm{p}|\rangle$ levels with Rabi frequency $\Omega_\mathrm{p}$. $n(z)$ is the atomic density of the ensemble. The transition operators $\hat{\sigma}_{mn}(z) = \ket{m(z)} \bra{n(z)}$ correspond to the operators for the transition of the atom located at point $z$ between states $\ket{m}$ and $\ket{n}$, $\{m,n\}\in\{g,e_\mathrm{p},r_\mathrm{p},e_\mathrm{c},r_\mathrm{c}\}$. Here, the rotating wave approximation was performed and the excitation regime corresponds to EIT conditions.

The Hamiltonian for the control branch under the rotating wave approximation reads
\begin{align}
\notag
\hat{\mathcal{H}}_{\mathrm{control}}=&\hbar\int\limits_{0}^{L} dzn(z)[ \Delta\hat{\sigma}_{e_\mathrm{c}e_\mathrm{c}}(z,t) +\delta\hat{\sigma}_{r_\mathrm{c}r_\mathrm{c}}(z,t)  
\\&- \Omega_\mathrm{c}\hat{\sigma}_{r_\mathrm{c}e_\mathrm{c}}(z,t)-\Omega_\mathrm{c}^*\hat{\sigma}_{e_\mathrm{c}r_\mathrm{c}}(z,t)   \notag
\\&+ (g_\mathrm{c} \hat{\mathcal{E}}_{c}\hat{\sigma}_{e_\mathrm{c}g}(z,t)  +g_\mathrm{c}^*\hat{\mathcal{E}}^{\dagger}_\mathrm{c} \hat{\sigma}_{ge_\mathrm{c}}(z,t)  )  ],
\label{eq:H_controlfs}
\end{align}
with detuning $\Delta = \omega_{e_\mathrm{c}g} - \omega_\mathrm{1,c}$, where $\omega_\mathrm{1,c}$ corresponds to the single photon wavepacket central frequency and the two-photon detuning is $\delta = \Delta - (-\omega_{r_\mathrm{c}e_\mathrm{c}} + \omega_\mathrm{2,c})$ as above. The single control photon is described by the operator $\hat{\mathcal{E}}^{\dagger}_\mathrm{c}$, which couples to each atom with strength $g_\mathrm{c}$, and the classical drive for the upper transition has Rabi frequency $\Omega_\mathrm{c}$.

The van der Waals interaction Hamiltonian for the ensemble can be written as
\begin{align}
\label{eq:H_Rydfs}
\hat{\mathcal{H}}_{\mathrm{Ryd}} =& \hbar \int_0^L dzn(z)\int_0^L dz'n(z')\mathcal{V}(z,z')\times\nonumber\\&\times\left( \hat{\sigma}_{r_\mathrm{p}r_\mathrm{p}}(z,t) \otimes \hat{\sigma}_{r_\mathrm{c}r_\mathrm{c}}(z',t)\right),
\end{align}
where $\mathcal{V}(z,z')=C_6/|z-z'|^6$ and $C_6$ is an interaction coefficient \cite{Saffman2010}. Similar to the cavity case, we consider a single incident control photon and allow multiple probe photons, but only include interactions between Rydberg atoms in different branches. This is justified when the principal quantum number of the control Rydberg state $\ket{r_\mathrm{c}}$ is significantly higher than that of the probe Rydberg state $\ket{r_\mathrm{p}}$, or when the probe and control Rydberg states are near a Förster resonance, as discussed in detail in Appendix \ref{appendix:a}. Lastly, we include the propagation $\hat{\mathcal{H}}_{\mathrm{propag}}$ and the reservoir $\hat{\mathcal{H}}_{\mathrm{res}}$ Hamiltonians.

In the following we consider the dynamics of the probe and control branches separately, by assuming that the characteristic timescales for probing are much faster than for storage of the control photon.


\subsection{\label{sec:fsprob}Probe-induced dephasing}
We now focus on the branch where the continuous probe field is incident [left branch in Fig. \ref{fig:sketch}(b)]. In this section we study the modification of the scattering properties of the probe field conditioned on the absence or presence of a control photon, which is the main function of the SPT \cite{Chang2007}. Furthermore we describe the effective dephasing processes induced by the continuous probe field on the stored control photon via the Rydberg interaction.
\par The dynamics of the probe branch are described by the equations of motion derived from the Hamiltonians $\hat{\mathcal{H}}_\mathrm{probe}+\hat{\mathcal{H}}_\mathrm{Ryd}$, the Lindblad jump operator (\ref{eq:gjum})
which accounts for the losses to
the environment of the $l$th atom on the probe excited state $\ket{e_\mathrm{p}}$ through spontaneous emission with a decay rate of  $\gamma_{e_\mathrm{p}}$
and the Maxwell equation for light propagation in the medium. Reabsorption of emitted photons is disregarded by presuming an elongated atomic ensemble along the $z$-axis, thereby minimizing the optical depth in the transverse direction. The EOMs for the operators associated with the probe branch are found using the Heisenberg equation (\ref{eq:heom}) to be
\begin{align}
&(\partial_t+c\partial_z)\hat{\mathcal{E}}_\mathrm{p}(z,t) =ig_\mathrm{p}n(z)L\hat{\sigma}_{ge_\mathrm{p}}(z,t), \label{eq:fsa_dot}
\\&\partial_t\hat{\sigma}_{ge_\mathrm{p}}(z,t) = -\gamma_{e_\mathrm{p}}\hat{\sigma}_{ge_\mathrm{p}}(z,t) + ig_\mathrm{p}^*\hat{\mathcal{E}}_\mathrm{p}(z,t) 
\notag
\\& \qquad\qquad\qquad + i \Omega_\mathrm{p} \hat{\sigma}_{gr_\mathrm{p}}(z,t) ,\label{eq:fssigma_ge_dot}
\\&
\notag
\partial_t\hat{\sigma}_{gr_\mathrm{p}}(z,t) = -i\int_0^Ldz'n(z')\mathcal{V}(z,z')\hat{\sigma}_{r_\mathrm{c}r_\mathrm{c}}(z',t)\otimes\hat{\sigma}_{gr_\mathrm{p}}(z,t) 
\\&\qquad\qquad\qquad+ i\Omega_\mathrm{p}^* \hat{\sigma}_{ge_\mathrm{p}}(z,t) .
\label{eq:fssigma_gr_dot}
\end{align}

Moving to a comoving frame under the transformation $t'=t-z/c$ with $c$ being speed of light in the medium, absorbing a factor $\sqrt{c/L\gamma_{e_\mathrm{p}}}$ in the definition of
$\hat{\mathcal{E}}_\mathrm{p}$ and assuming $g_\mathrm{p}$ real without loss of generality, we define the optical depth of a single atom for the probe branch as $d_\mathrm{p1}=g^2_\mathrm{p}L/c \gamma_{e_\mathrm{p}}$.
Dropping the primes on the time coordinate, the EOMs of
the operators associated with the probe branch are then
found to be
\begin{align}
&\partial_{z}\hat{\mathcal{E}}_\mathrm{p}(z,t)  =i\sqrt{d_\mathrm{p1}}n(z)\hat{\sigma}_{ge_\mathrm{p}}(z,t), \label{eq:fsa_dot1}
\\&\partial_t\hat{\sigma}_{ge_\mathrm{p}}(z,t) = -\gamma_{e_\mathrm{p}}\hat{\sigma}_{ge_\mathrm{p}}(z,t) + i \sqrt{d_\mathrm{p1}}\gamma_{e_\mathrm{p}} \hat{\mathcal{E}}_\mathrm{p}(z,t)\notag 
\\&\label{eq:fssigma_ge_dot1} \qquad\qquad\qquad+ i \Omega_\mathrm{p} \hat{\sigma}_{gr_\mathrm{p}}(z,t) ,\\
&\partial_t\hat{\sigma}_{gr_\mathrm{p}}(z,t) = -i\int^L_0dz'n(z')\mathcal{V}(z,z')\hat{\sigma}_{r_\mathrm{c}r_\mathrm{c}}(z',t)\otimes\hat{\sigma}_{gr_\mathrm{p}}(z,t)\notag
\\& \qquad\qquad\qquad+ i\Omega_\mathrm{p}^* \hat{\sigma}_{ge_\mathrm{p}}(z,t) \label{eq:fssigma_gr_dot1},
\end{align}
\par This system of three EOMs can be reduced to two by formally integrating the probe field equation (\ref{eq:fsa_dot1}) and substituting it into Eq. (\ref{eq:fssigma_ge_dot1}). This gives
\begin{align}
&\partial_t\hat{\sigma}_{ge_\mathrm{p}}(z,t) = -\gamma_{e_\mathrm{p}}\hat{\sigma}_{ge_\mathrm{p}}(z,t) - d_\mathrm{p1}\gamma_{e_\mathrm{p}}\int^z_0dz'n(z')\hat{\sigma}_{ge_\mathrm{p}}(z',t) \notag
\\& \qquad\qquad\qquad +i \Omega_\mathrm{p} \hat{\sigma}_{gr_\mathrm{p}}(z,t)
+i\sqrt{d_\mathrm{p1}}\gamma_{e_\mathrm{p}} \hat{\mathcal{E}}_\mathrm{p}(0,t).
\end{align}
We notice that the propagating probe field mediates an effective coupling between the atoms due to the collective optical response of the ensemble. The effective Lindblad operator associated with the atomic effective coupling by propagation through the optically dense ensemble is
\begin{align}
&\hat{L}_{1d_\mathrm{p}}(z)=\sqrt{2d_\mathrm{p1}\gamma_{e_\mathrm{p}}}\int^{z}_0dz'n(z')\hat{\sigma}_{ge_\mathrm{p}}(z').
\end{align}
By integrating over the entire length of the ensemble, we obtain the Lindblad jump operator describing the loss of a probe photon by transmission through the entire ensemble. This is given by

\begin{equation}
 \hat{L}_{d_\mathrm{p}}=\int^L_0dz\hat{L}_{1d_\mathrm{p}}(z)=\sqrt{2d_\mathrm{p1}\gamma_{e_\mathrm{p}}}\int^L_0dzn(z)\hat{\sigma}_{ge_\mathrm{p}}(z). \label{eq:fsdjum}
\end{equation}

As for the cavity case, we consider the probe input field  to be prepared in a coherent state, consisting of a classical amplitude $\alpha_\mathrm{in,p}$ and vacuum fluctuations, i.e., $\hat{\mathcal{E}}_\mathrm{p}(z=0)=\alpha_\mathrm{in,p}\sqrt{2/\gamma_{e_\mathrm{p}}}+\delta\hat{\mathcal{E}}_\mathrm{p}(z=0)$. Similarly to our approach for the cavity case in Sec. \ref{sec:probcav},  we solve the system in the absence of a stored excitation in the control branch, which we use as a reference, the subtraction of which leads to a purely Rydberg-interaction-related description. In the case of no stored excitation in the control branch, the EOM of the transition operator to the probe Rydberg state (\ref{eq:fssigma_gr_dot1}) simplifies to
\begin{equation}
\partial_t\hat{\sigma}_{g{r_\mathrm{p}}}(z,t)=i\Omega^*_\mathrm{p}\hat{\sigma}_{ge_\mathrm{p}}(z,t).\label{eq:fssigma_gr_dot12}
\end{equation}
Moving to the frequency domain under the Fourier transform (\ref{eq:four}) where $\omega$ is defined with respect to the central frequency of the photon, we can solve the set of EOMs (\ref{eq:fsa_dot1}), (\ref{eq:fssigma_ge_dot1}), (\ref{eq:fssigma_gr_dot12}). The solutions have the form $\hat{o}=\overline{o}+\mathcal{O}(\delta\hat{\mathcal{E}}_\mathrm{p}(z=0))$ and are comprised of a classical part proportional to the probe field amplitude $\alpha_\mathrm{in,p}$ and a quantum part proportional to the fluctuation operator $\delta\hat{\mathcal{E}}_\mathrm{p}(z=0)$.
The classical parts of the solutions are then
\begin{align}
&\overline{\mathcal{E}}_\mathrm{p}(z,\omega)=e^{-\frac{\int_0^zdz'n(z')\gamma_{e_\mathrm{p}} d_\mathrm{p1}  (-i\omega)}{(\gamma_{e_\mathrm{p}}-i\omega)(-i\omega)+|\Omega_\mathrm{p}|^2}}\alpha_\mathrm{in,p}(\omega)\sqrt{2/\gamma_{e_\mathrm{p}}}, \label{eq:fsovere}
\\&
\overline{\sigma}_{ge_{\mathrm{p}}}(z,\omega) =\frac{ i\sqrt{2d_\mathrm{p1}\gamma_{e_\mathrm{p}}}(-i\omega)\alpha_\mathrm{in,p}(\omega)e^{-\frac{\int_0^zdz'n(z')\gamma_{e_\mathrm{p}} d_\mathrm{p1} (-i\omega)}{(\gamma_{e_\mathrm{p}}-i\omega)(-i\omega)+|\Omega_\mathrm{p}|^2}}}{(\gamma_{e_\mathrm{p}}-i\omega)(  -i\omega)+|\Omega_\mathrm{p}|^2},\label{eq:fsoverseg}
\\&
\overline{\sigma}_{gr_{\mathrm{p}}}(z,\omega) =\frac{- \sqrt{2d_\mathrm{p1}\gamma_{e_\mathrm{p}}}\Omega^*_\mathrm{p}\alpha_\mathrm{in,p}(\omega)e^{-\frac{\int_0^zdz'n(z')\gamma_{e_\mathrm{p}} d_\mathrm{p1}  (-i\omega)}{(\gamma_{e_\mathrm{p}}-i\omega)(-i\omega)+|\Omega_\mathrm{p}|^2}}}{(\gamma_{e_\mathrm{p}}-i\omega)(  -i\omega)+|\Omega_\mathrm{p}|^2}.\label{eq:fsoversrg}
\end{align}
\par The measure of the scattering properties of the system is the transmission coefficient, described by the relation $\hat{\mathcal{E}}_\mathrm{p}[L,\omega]=T_\mathrm{p}[\omega]\hat{\mathcal{E}}_\mathrm{p}[0,\omega]$. Using Eq. (\ref{eq:fsovere}), the transmission coefficient in the absence of control excitation is 
\begin{equation}
T_\mathrm{p}[\omega]=e^{\frac{i\omega\gamma_{e_\mathrm{p}} d_\mathrm{p1}N  }{(\gamma_{e_\mathrm{p}}-i\omega)(-i\omega)+|\Omega_\mathrm{p}|^2}}.    \label{eq:tr}
\end{equation}
Since the device is operating under EIT conditions in the absence of control excitation, we note that on resonance the probe field is fully transmitted $T_\mathrm{p}[\omega=0]=1$.
\par Furthermore, to achieve a description solely of the Rydberg-associated processes, the solutions of the probe branch's EOMs in the absence of any stored control excitation, (\ref{eq:fsovere})-(\ref{eq:fsoversrg}) are used as a reference, and a new set of shifted operators is introduced by subtracting the reference frame from the original operators. The shifted operators related to the probe branch are then introduced as
\begin{align}
&\delta\hat{\mathcal{E}}_\mathrm{p}(z,\omega)=\hat{\mathcal{E}}_\mathrm{p}(z,\omega)-\overline{\mathcal{E}}_\mathrm{p}(z,\omega),\label{eq:de1}
\\
&
\delta\hat{\sigma}_{ge_\mathrm{p}}(z,\omega)=\hat{\sigma}_{ge_{\mathrm{p}}}(z,\omega) -\overline{\sigma}_{ge,\mathrm{p}}(z,\omega),\\
&
\delta\hat{\sigma}_{gr_\mathrm{p}}(z,\omega)=\hat{\sigma}_{gr_{\mathrm{p}}}(z,\omega) -\overline{\sigma}_{gr,\mathrm{p}}(z,\omega)\label{eq:de3},
\\&\delta\hat{\mathcal{E}}_\mathrm{p}(z=0,\omega)=\hat{\mathcal{E}}_\mathrm{p}(z=0,\omega)-\alpha_\mathrm{in,p}(\omega)\sqrt{2/\gamma_{e_\mathrm{p}}}.
\end{align}
The EOMs for the shifted operators are derived by substituting the definitions of the shifted operators (\ref{eq:de1})--(\ref{eq:de3}) into the EOMs of the probe branch (\ref{eq:fsa_dot1})--(\ref{eq:fssigma_gr_dot1}), accounting for the possible presence of a control photon excitation in the control branch. The EOMs for the shifted operators, associated with the probe branch, in the frequency domain are 
\begin{align}
&\partial_z\delta\hat{\mathcal{E}}_\mathrm{p}(z,\omega)=i\sqrt{d_\mathrm{p1}}n(z)\delta\hat{\sigma}_{ge_\mathrm{p}}(z,\omega),\\
&\nonumber
-i\omega\delta\hat{\sigma}_{ge_\mathrm{p}}(z,\omega)=-\gamma_{e_\mathrm{p}}\delta\hat{\sigma}_{ge_\mathrm{p}}(z,\omega)
+i\Omega_\mathrm{p}\delta\hat{\sigma}_{gr_\mathrm{p}}(z,\omega)
\\&\qquad\qquad\qquad\qquad+i\sqrt{d_\mathrm{p1}}\gamma_{e_\mathrm{p}}\delta\hat{\mathcal{E}}_\mathrm{p}(z,\omega),\\
&
-i\omega\delta\hat{\sigma}_{gr_\mathrm{p}}(z,\omega)=+i\Omega^*_\mathrm{p}\delta\hat{\sigma}_{ge_\mathrm{p}}(z,\omega)
\nonumber
\\&-i\int^L_0dz'n(z')\mathcal{V}(z,z')\hat{\sigma}_{r_\mathrm{c}r_\mathrm{c}}(z',\omega)\otimes\delta\hat{\sigma}_{gr_\mathrm{p}}(z,\omega)\nonumber
\\&+\frac{i\sqrt{2d_\mathrm{p1}\gamma_{e_\mathrm{p}}}\Omega_\mathrm{p}^*\alpha_\mathrm{in}\int^L_0dz'n(z')\mathcal{V}(z,z')\hat{\sigma}_{r_\mathrm{c}r_\mathrm{c}}(z',\omega)}{(-i\omega)(\gamma_{e_\mathrm{p}}-i\omega)+|\Omega_\mathrm{p}|^2}.\label{eq:fssgr}
\end{align}
As observed, a classical feeding term appears in Eq. (\ref{eq:fssgr}) only in the presence of a stored excitation in the control branch, resulting from the Rydberg-Rydberg interaction between excitations in the probe and control branches.
\par The solutions of these EOMs are expressed as $\hat{o}[\omega]=\overline{o}[\omega]\hat{\sigma}_{rr}+\mathcal{O}(\delta\hat{\mathcal{E}}_\mathrm{p}[0,\omega])$, where the first term represents a classical component dependent on the probe field's amplitude $\alpha_\mathrm{in,p}$, conditioned on the presence of a stored control excitation. The second term corresponds to a weak quantum component proportional to the fluctuation operator $\delta\hat{\mathcal{E}}_\mathrm{p}[0,\omega]$. Neglecting the weak second term, in a similar fashion as for the cavity case, the solution for the probe excited-state transition operator, for a resonant probe field (i.e., $\omega = 0$, a field constant in time), is found to be
\begin{align}
&
\delta\hat{\sigma}_{ge_\mathrm{p}}(z)=-\frac{i\sqrt{2}\alpha_\mathrm{in,p}}{\sqrt{d_\mathrm{p1}\gamma_{e_\mathrm{p}}}}\int^L_0dz'n(z')\hat{\sigma}_{r_\mathrm{c}r_\mathrm{c}}(z')d_\mathrm{b1,p}(z,z')\times \nonumber
\\&\times\left(\int_0^zdz''n(z'')d_\mathrm{b1,p}(z'',z')e^{-\int^z_{z''}dz'''n(z''')d_\mathrm{b1,p}(z''',z')}-1\right).\label{eq:fsge3}
\end{align}
The single-atom blockaded optical depth is defined as the optical depth of an atom located at position $z$ due to the control excitation stored in the Rydberg state $\ket{r_\mathrm{c}}$ of the atom located at position $z'$
\begin{equation}
d_\mathrm{b1,p}(z,z')=\frac{d_\mathrm{p1}\gamma_{e_\mathrm{p}}}{\gamma_{e_\mathrm{p}}+\frac{|\Omega_\mathrm{p}|^2}{i\mathcal{V}(z,z')}}.
\end{equation}
Additionally, the blockaded optical depth of the probe branch due to a stored control excitation in Rydberg state $\ket{r_\mathrm{c}}$ of the atom at position $z'$ is given by $d_\mathrm{b,p}(z')=\int_0^Ldzd_\mathrm{b1,p}(z,z')$.
\par At this stage, the primary function of the SPT becomes evident. When a control excitation is stored at the Rydberg state $\ket{r_\mathrm{c}}$ of the atom positioned at point $z'$, the scattering properties of the probe field deviate from the full transmission described by Eq. (\ref{eq:tr}). This effect is confirmed by the solution for the probe field in the presence of a stored control excitation, where the resonance transmission coefficient is given by
\begin{equation}\label{eq:trk}
T_\mathrm{p}[z',\omega=0]=e^{-d_\mathrm{b,p}(z')},
\end{equation}
where $z'$ represents the location where the control photon is stored.

Equation (\ref{eq:trk}) can be interpreted as an effective two-level system sphere centered at point $z'$ with a radius equal to the Rydberg blockade radius, residing within the ensemble of three-level atoms. As is evident from Eq. (\ref{eq:tr}) and (\ref{eq:trk}) the presence of stored excitation modifies the scattering properties of the system, by an amount set by the blockaded optical depth $d_\mathrm{b,p}(z')$. This modification leads in return to information regarding the presence and location of the stored control excitation.

This takes us to the second part of the section, relating to how the extraction of information on the presence or location of the stored excitation in the control branch leads to dephasing processes. These processes arise from the two Lindblad operators (\ref{eq:fsdjum}) and the equivalent of (\ref{eq:gjum}). At this point we note that we can move to a discrete description of our system by dividing the system into pieces, each corresponding to a single atom at position $z$. To do so, the following transformations are used $\hat{\sigma}(z_l,t)\rightarrow\hat{\sigma}^l(t)$, $\int^{z_{l'}}_{z_l}dzn(z)\rightarrow\sum_{i=l}^{l'}$. Subsequently the discrete version of Eqs. (\ref{eq:gjum}) and (\ref{eq:fsdjum}) in the shifted frame using the solution (\ref{eq:fsge3}) read
\begin{align}
&\hat{L}_{d_\mathrm{p}}=\sqrt{d_\mathrm{p1}}\sum_{k=1}^N\sum_{\substack{l=1\\ l\neq k}}^N\sqrt{\tilde{\gamma}_{ge_\mathrm{p}}^{k,l}}\hat{\sigma}_{r_\mathrm{c}r_\mathrm{c}}^k,\label{eq:djump}
\\&\hat{L}_{ge_\mathrm{p}}^l=\sum_{\substack{k=1\\ k\neq l}}^N\sqrt{\tilde{\gamma}_{ge_\mathrm{p}}^{k,l}}\hat{\sigma}_{r_\mathrm{c}r_\mathrm{c}}^k,\label{eq:ejump}
\end{align}
where the dephasing rate of the Rydberg state of atom $k$ due to the decay of atom $l$ is given by
\begin{align}
&\sqrt{\tilde{\gamma}_{ge_\mathrm{p}}^{k,l}}= -i\frac{2}{\sqrt{d_\mathrm{p1}}}d_\mathrm{b1,p}^{k,l}\alpha_\mathrm{in,p}D_{b,p}^{k,l},\label{eq:fserate}
\end{align}
and $D_\mathrm{b,p}^{k,l}=\sum_{l'=1}^ld^{k,l'}_\mathrm{b1,p}e^{-\sum_{l''=l'}^ld^{k,l''}_\mathrm{b1,p}}-1$ is the blockaded optical depth attenuation due to propagation of the probe in the atomic medium past the $l$th atom due to the Rydberg excitation of $k$th atom.

These operators represent the dephasing of control excitations due a change in the transmission and the dephasing due to the spontaneous emission from the probe excited state $\ket{e_\mathrm{p}}$ of the $l$th atom respectively. The dephasing rates on the $k$th atom, associated with the two processes, are given by $\gamma_i^k=\bra{r^k_\mathrm{c}}\frac{1}{2}\hat{L}_{i}^\dagger\hat{L}_{i}\ket{r^k_\mathrm{c}}/\braket{r_\mathrm{c}^k}$ and read
\begin{align}
&\gamma_{d_\mathrm{p}}^k\label{eq:drate}
=\frac{d_\mathrm{p1}}{2}\left|\sum^N_{\substack{l=1\\ l\neq k}}\sqrt{\tilde{\gamma}_{ge_\mathrm{p}}^{k,l}}\right|^2=2\left|\sum^N_{\substack{l=1\\ l\neq k}} d_\mathrm{b1,p}^{k,l}D_\mathrm{b,p}^{k,l}\alpha_\mathrm{in,p}\right|^2,
\\&\gamma_{ge_\mathrm{p}}^{k,l}=\frac{1}{2}\left|\sqrt{\tilde{\gamma}_{ge_\mathrm{p}}^{k,l}}\right|^2=\sum^N_{\substack{k=1\\ k\neq l}}\frac{2|\alpha_\mathrm{in,p}|^2}{d_\mathrm{p1}}\left|d_\mathrm{b1,p}^{k,l}D_\mathrm{b,p}^{k,l}\right|^2.
\end{align}
\par Let us first consider the dephasing operator $\hat{L}_{ge_\mathrm{p}}^l$ due to spontaneous emission of the $l$th atom from the probe excited state $\ket{e_\mathrm{p}}$. Similar to the description in Sec. \ref{sec:probcav}, this operator depends on the specific atom that has undergone decay, and thus imparts information to the environment regarding the collective Rydberg excitation's spatial position. This results in the localization of the stored control excitation around the decayed atom. This localization extends the lifetime of the stored control excitation (see Sec. \ref{sec:resfs}).
\par The second dephasing operator $\hat{L}_{d_\mathrm{p}}$ acts on the stored excitation due to a change in transmission. This dephasing process reveals information about the presence of the stored excitation via the change of the transmitted probe field, which can be detected. The transmission change becomes significant for large values of the blockaded optical depth, i.e., $d_\mathrm{b,p}\gg 1$, as seen in Eq. (\ref{eq:drate}). This is the detectable process that allows the readout of the signal (see Sec. \ref{sec:resfs}). Unlike the cavity case, this operator also leads to localization, since the storage occurs in an exponentially decaying mode due to the decreasing probability of an excitation propagating through the ensemble without being dephased.
\par 
Lastly, we define the total dephasing rate on the $k$th atom of the control stored Rydberg excitation resulting from the Rydberg-mediated decay processes in the probe branch as
\begin{equation}\label{eq:rater}
\gamma_{r}^k=\gamma^k_{d_\mathrm{p}}+\sum^N_{\substack{l=1\\ l\neq k}}\gamma_{ge_\mathrm{p}}^{k,l}.
\end{equation}
The average total dephasing rate $\overline{\gamma}_{r}=\frac{1}{N}\sum_{k=1}^N\gamma_{r}^k$ is proportional to the strength of the probing field $|\alpha_\mathrm{in,p}|^2$. Consequently, we can optimize the impedance matching conditions for control-photon storage by appropriately adjusting the strength of the probing field $|\alpha_\mathrm{in,p}|^2$, as discussed in the following section.


\subsection{\label{sec:fsim}Impedance matching}
A crucial aspect of the SPT protocol, as also demonstrated for the case of the cavity, is the efficient conversion of an incident control photon into a Rydberg collective excitation with a high probability. To accomplish this, we examine the scattering dynamics occurring in the control branch and derive an analytical estimate for the optimal impedance-matching conditions.
\par The dynamics of the control branch are described by the EOMs for the corresponding operators derived from $\hat{H}_\mathrm{control}$, the Maxwell equation for light propagation in the medium and the Lindblad jump operators 
\begin{align}
&\hat{L}_{ge_\mathrm{c}}(z)=\sqrt{2\gamma_{e_\mathrm{c}}}\hat{\sigma}_{ge_\mathrm{c}}(z),\label{eq:fsgjum}
\\&\hat{L}_{r_\mathrm{c}r_\mathrm{c}}(z)=\sqrt{2\gamma_{r}}\hat{\sigma}_{r_\mathrm{c}r_\mathrm{c}}(z).\label{eq:lrc}
\end{align}
 The first Lindblad operator accounts for the losses to the environment via spontaneous emission from the control excited state $\ket{e_\mathrm{c}}$ of the atom located at point $z$. The second Lindblad operator (\ref{eq:lrc}) characterizes the total effective dephasing of the control Rydberg state $\ket{r_\mathrm{c}}$, induced by the decay effects of the probe branch's Lindblad operators (\ref{eq:djump}) and(\ref{eq:ejump}) mediated by the Rydberg interaction, derived in Sec. \ref{sec:fsprob}. In addition, the Lindblad jump operator describing the loss of a control photon via transmission through the ensemble, analogous to Eq. ($\ref{eq:fsdjum}$) for probe photons, is given by

\begin{equation}
 \hat{L}_{d_\mathrm{c}}=\sqrt{2d_\mathrm{c1}\gamma_{e_\mathrm{c}}}\int^L_0dzn(z)\hat{\sigma}_{ge_\mathrm{c}}(z). \label{eq:fsdpjum}
\end{equation}
\par The EOMs for the operators associated with the control branch are derived using the Heisenberg equation (\ref{eq:heom}) and read
\begin{align}
\label{eq:a_dot}
&(\partial_t+c\partial_z)\hat{\mathcal{E}}_{\mathrm{c}}(z,t) =ig_\mathrm{c}n(z)L\hat{\sigma}_{ge_\mathrm{c}}(z,t),\\
\label{eq:sigma_ge_dot}
&\partial_t\hat{\sigma}_{ge_\mathrm{c}}(z,t) = -(\gamma_{e_\mathrm{c}} + i \Delta) \hat{\sigma}_{ge_{\mathrm{c}}}(z,t) + i g_{\mathrm{c}}^* \hat{\mathcal{E}}_{\mathrm{c}}(z,t) \nonumber
\\&\qquad\qquad\qquad+ i \Omega_{\mathrm{c}} \hat{\sigma}_{gr_{\mathrm{c}}}(z,t) ,\\
\label{eq:sigma_gr_dot}
&\partial_t\hat{\sigma}_{gr_{\mathrm{c}}}(z,t) = -\Big(\gamma_r + i\delta  \Big) \hat{\sigma}_{gr_{\mathrm{c}}}(z,t) + i\Omega_{\mathrm{c}}^* \hat{\sigma}_{ge_{\mathrm{c}}}(z,t) .
\end{align}

Moving to a comoving frame under the transformation $t'=t-z/c$, absorbing a factor $\sqrt{c/L\gamma_{e_\mathrm{c}}}$ in the definition of
$\hat{\mathcal{E}}_\mathrm{c}$ and assuming $g_\mathrm{c}$ real without loss of generality, we define the optical depth of a single atom for the control branch as $d_\mathrm{c1}=g_\mathrm{c}^2L/c \gamma_{e_\mathrm{c}}$.
Dropping the primes on the time coordinate, the EOMs of
the operators associated with the control branch are then
found to be
\begin{align}
\label{eq:a_dot}
&\partial_{z}\hat{\mathcal{E}}_{\mathrm{c}}(z,t) =i\sqrt{d_\mathrm{c1}}n(z)\hat{\sigma}_{ge_\mathrm{c}}(z,t),\\
\label{eq:sigma_ge_dot}
&\partial_t\hat{\sigma}_{ge_\mathrm{c}}(z,t) = -(\gamma_{e_\mathrm{c}}+i\Delta)\hat{\sigma}_{ge_{\mathrm{c}}}(z,t) \nonumber
\\&\qquad\qquad\qquad+ i \sqrt{d_\mathrm{c1}}\gamma_{e_\mathrm{c}} \hat{\mathcal{E}}_{\mathrm{c}}(z,t) + i \Omega_{\mathrm{c}} \hat{\sigma}_{gr_{\mathrm{c}}}(z,t) ,\\
\label{eq:sigma_gr_dot}
&\partial_t\hat{\sigma}_{gr_{\mathrm{c}}}(z,t) = -(\gamma_r+i\delta)\hat{\sigma}_{gr_\mathrm{c}}(z,t) + i\Omega_{\mathrm{c}}^* \hat{\sigma}_{ge_{\mathrm{c}}}(z,t) .
\end{align} 
\par Subsequently, we move to the frequency domain under the Fourier transform (\ref{eq:four}).
Solving these EOMs of the control branch, we obtain the transmission coefficient and the susceptibilities corresponding to transitions from the ground state to the excited state $\ket{e_\mathrm{c}}$ and to the Rydberg state $\ket{r_\mathrm{c}}$
\begin{align}
    & T_\mathrm{c}[\omega]=\exp\left\{-\frac{\gamma_{e_\mathrm{c}} d_\mathrm{c}  (\gamma_r+i(\delta-\omega))}{\left[\gamma_{e_\mathrm{c}}+i(\Delta-\omega)\right]\left[\gamma_r + i(\delta -\omega)+|\Omega_\mathrm{c}|^2\right]}\right\}  ,\label{eq:tc}
    \\&   \chi_{e_\mathrm{c}}[\omega]= \frac{ i\sqrt{d_\mathrm{c1}}(\gamma_r + i(\delta -\omega))}{\left[\gamma_{e_\mathrm{c}}+i(\Delta-\omega)\right]\left[\gamma_r + i(\delta -\omega)+|\Omega_\mathrm{c}|^2\right]}T_\mathrm{c}[\omega],
    \\&\chi_{r_\mathrm{c}}[\omega]= \frac{ -\sqrt{d_\mathrm{c1}}\Omega^*}{\left[\gamma_{e_\mathrm{c}}+i(\Delta-\omega)\right]\left[\gamma_r + i(\delta -\omega)+|\Omega_\mathrm{c}|^2\right]}T_\mathrm{c}[\omega],\label{eq:rcc}
\end{align}
respectively. These proportionality factors relate the control branch's operators to the control input field through $\hat{\mathcal{E}}_\mathrm{c}(L)=T_\mathrm{c}\hat{\mathcal{E}}_\mathrm{c}(0)$,  $\int^L_0dz\hat{\sigma}_{ge_\mathrm{c}}=\chi_{e_\mathrm{c}} \hat{\mathcal{E}}_\mathrm{c}(0)$ and  $\int^L_0dz\hat{\sigma}_{gr_\mathrm{c}}=\chi_{r_\mathrm{c}}\hat{\mathcal{E}}_\mathrm{c}(0)$.
\par Conservation of probability dictates the balancing of the incoming and outgoing scattering processes of the system. For the scattering effects of the control branch, this reads
\begin{align}
\notag
\langle (\hat{\mathcal{E}}_\mathrm{c}(0))^\dagger\hat{\mathcal{E}}_\mathrm{c}(0)\rangle=&\langle (\hat{\mathcal{E}}_\mathrm{c}(L))^\dagger\hat{\mathcal{E}}_\mathrm{c}(L)\rangle+\int^L_0dz\frac{\langle\hat{L}_{ge_\mathrm{c}}^\dagger(z) \hat{L}_{ge_\mathrm{c}}(z)\rangle}{\gamma_{e_\mathrm{c}}}
\\&+\frac{1}{\gamma_{e_\mathrm{c}}}\int^L_0dz\langle \hat{L}^\dagger_{r_\mathrm{c}}(z) \hat{L}_{r_\mathrm{c}}(z)\rangle.\label{eq:1probfs}
\end{align}
By the use of the proportionality factors (\ref{eq:tc})--(\ref{eq:rcc}), which are independent of the spatial coordinate $z$, Eq. (\ref{eq:1probfs}) can be expressed as
\begin{equation}
|T_\mathrm{c}[\omega]|^2+\Gamma_{e_\mathrm{c}}[\omega]+\Gamma_{r_\mathrm{c}}[\omega]=1,\label{eq:probfs}
\end{equation}
where we have introduced the loss probability via spontaneous emission of the control excited state $\ket{e_\mathrm{c}}$ for an incident photon $\Gamma_{e_\mathrm{c}}[\omega]=2|\chi_{e_\mathrm{c}}[\omega]|^2$,  the dephasing probability of the Rydberg control state $\ket{r_\mathrm{c}}$ of an incident photon $\Gamma_{r_\mathrm{c}}[\omega]=2\gamma_r|\chi_{r_\mathrm{c}}[\omega]|^2/\gamma_{e_\mathrm{c}}$ and the transmittance of the incoming control field $|T_\mathrm{c}[\omega]|^2$. Equation (\ref{eq:probfs}) naturally reflects the expected behavior, as it simply expresses the conservation of probability.
\par To be impedance matched, the dephasing probability of the Rydberg control state $\Gamma_{r_\mathrm{c}}$ should be close to unity and accordingly the transmittance and the decay rate through the excited state should be close to zero.
In the same way as for the cavity case, we adjust the probe field strength so that the dephasing rate is equal to the decay rate of a fully delocalized Rydberg state, which represent the maximum speed at which excitations can enter the system,
\begin{equation}
    \gamma_r=d_\mathrm{c}\gamma_{e_\mathrm{c}}\Omega_\mathrm{c}^2/\Delta^2,\label{eq:gopt}
\end{equation}
where $d_\mathrm{c}=d_\mathrm{c1}N$ is the total optical depth of the control branch \cite{PhysRevLett.100.093603}.
Furthermore, similar to the cavity case, the detuning is chosen to be large compared with the effective decay rate from the control excited stated, i.e., $\Delta\gg d_\mathrm{c}\gamma_{e_\mathrm{c}}$. 
Under this condition the effective escape rate at the end of the ensemble is proportional to the number of atoms participating in the collective excitation \cite{Gorshkov2007b,Zeuthen2017}, allowing localization to suppress this type of decay. Finally, in order to account for the ac Stark shift, which is present due to the control branch's driving field $\Omega_\mathrm{c}$, the two photon detuning is set to be equal to the ac Stark shift, i.e., 
$\delta=|\Omega_\mathrm{c}|^2/\Delta$.
\par Under these conditions the transmittance on resonance is found to be
\begin{equation}
|T_\mathrm{c}[\omega=0]|^2=\exp\left\{-2\frac{\frac{\gamma_e^2d_\mathrm{c}^3}{\Delta^2}+d_\mathrm{c}^2+d_\mathrm{c}}{(1+d_\mathrm{c})^2}\right\},
\end{equation}
which goes to zero for large values of the optical depth, i.e., $d_\mathrm{c}\gg1$.
Accordingly, the dephasing probability of the Rydberg state $\Gamma_{r_\mathrm{c}}$ goes to unity for large optical depth, i.e., $d_\mathrm{c}\gg 1$, since
\begin{equation}
\Gamma_{r_\mathrm{c}}[\omega=0]=\frac{d_\mathrm{c}}{1+d_\mathrm{c}}(1-|T_\mathrm{c}[\omega=0]|^2),\label{eq:imcfs}
\end{equation}
resulting in the fulfillment of the impedance-matching condition. Furthermore, the spontaneous emission loss probability $\Gamma_{e_\mathrm{c}}$ goes to zero in the limit of large optical depth $d_\mathrm{c}\gg 2$, since  
\begin{equation}
\Gamma_{e_\mathrm{c}}[\omega=0]=
\frac{1}{1+\frac{1}{d_\mathrm{c}}}(1-|T_\mathrm{c}[\omega=0]|^2).
\end{equation}
This analytical estimate is confirmed by numerical simulations discussed in Sec. \ref{sec:resfs}.


\subsection{\label{sec:simfs}Numerical simulation}
In the current section, we describe the wave-function Monte Carlo (wfMC) approach \cite{Dalibard1992,Molmer1993} used to simulate the system. The system is sketched in Fig.~\ref{fig:fssketch} and assumes the propagation of both probe and control signals in the Rydberg medium along the $z$ axis. The atomic medium consists of 1000 atoms randomly placed with a Gaussian distribution in 1D over the length $L$ along the $z$ axis. We keep the total optical depth $d_\mathrm{c}$ much less than the total number of atoms to represent a situation with weakly coupled atoms. 
The random neighboring distance between the atoms due to the Gaussian distribution leads to different values of the single atom blockaded optical depths $d_\mathrm{b,p}^k$ depending on the location of each atom. We thus describe the system using the average blockaded optical depth, defined as $\overline{d}_\mathrm{b,p}=\frac{1}{N}\sum^N_{k=1}d_\mathrm{b,p}^k$.

\par The system is described by the control Hamiltonian (\ref{eq:H_controlfs}), two decay operators (\ref{eq:fsdpjum}) and(\ref{eq:fsgjum}) and two dephasing operators (\ref{eq:djump}) and (\ref{eq:ejump}).
The two dephasing operators effectively account for the probe branch, which has been adiabatically eliminated using the results of Sec. \ref{sec:fsprob}. 
The non-Hermitian Hamiltonian of the full effectively described system in discrete form reads
\begin{align}\notag
\hat{\mathcal{H}}_{NH}=&\hat{\mathcal{H}}_{\mathrm{control}}-\frac{i}{2}\hat{L}_{d_\mathrm{c}}^\dagger\hat{L}_{d_\mathrm{c}}-\frac{i}{2}\sum_{l=1}^N(\hat{L}^l_{ge_\mathrm{c}})^\dagger\hat{L}^l_{ge_\mathrm{c}}
\\&-\frac{i}{2}\hat{L}_{d_\mathrm{p}}^\dagger\hat{L}_{d_\mathrm{p}}-\frac{i}{2}\sum_{l=1}^N(\hat{L}_{ge_\mathrm{p}}^l)^\dagger\hat{L}_{ge_\mathrm{p}}^l
\end{align}
The basis of the Hilbert space associated with the system is given by the $2N$-dimensional vector  $\{(\ket{e_\mathrm{c}^1},...,\ket{e_\mathrm{c}^N},\ket{r_\mathrm{c}^1},...,\ket{r_\mathrm{c}^N}\}$. The non-Hermitian Hamiltonian projected onto this basis can be written in matrix form as the $2N
\times2N$ matrix

\[
\bold{H_{NH}}=\left(
 \scalemath{0.68}{\begin{array}{cccccccc}
\Delta-\tilde{\gamma}_{e_\mathrm{c}}  & 0 & \cdots  & 0  & \Omega & 0 & \cdots& 0 \\
 -id_\mathrm{c1} \gamma_{e_\mathrm{c}} & \Delta -i \tilde{\gamma}_{e_\mathrm{c}} & 0  & \cdots & 0 & \Omega &\ddots& \vdots \\
\vdots & \ddots &  \ddots  & \ddots & \vdots  & \ddots & \ddots& 0 \\
-id_\mathrm{c1} \gamma_{e_\mathrm{c}} & \cdots & -id_\mathrm{c1} \gamma_{e_\mathrm{c}}  & \Delta -i \tilde{\gamma}_{e_\mathrm{c}} & 0  & \cdots & 0& \Omega \\
  \Omega & 0 & \cdots  & 0 & \delta - i\gamma_r^1 & 0 & \cdots& 0 \\
    0 & \Omega & 0  & \cdots & 0 & \delta - i\gamma_r^2 & \ddots& \vdots \\
\vdots & \ddots  &  \ddots  & \ddots & \vdots & \ddots & \ddots& 0 \\
0 & \cdots & 0 & \Omega  & 0 & \cdots & 0 & \delta - i \gamma_r^N\\
\end{array}}
\right)
\]
where $\gamma_r^k=\gamma_{d_\mathrm{p}}^k+\sum^N_{\substack{l=1\\ l\neq k}}\gamma_{ge_\mathrm{p}}^{k,l}$ is the effective dephasing rate resulting from decay processes on the probe branch (\ref{eq:djump}) and (\ref{eq:ejump}), and  $\tilde{\gamma}_{e_\mathrm{c}}=(1+d_\mathrm{c1}) \gamma_{e_\mathrm{c}}$ is the total excited-state decay.

\par In a similar manner as above, the wave-function describing the system at time $t$ is
\begin{equation}
\ket{\Psi(t)}=\sum_{l=1}^N c^l_{e_\mathrm{c}}(t)\ket{e_\mathrm{c}}_l+\sum_{l=1}^N c^l_{r_\mathrm{c}}(t)\ket{r_\mathrm{c}}_l.
\end{equation}
The system is not excited initially, i.e., $c^l_{e_\mathrm{c}}(t_0)=c^l_{r_\mathrm{c}}(t_0)=0$, for $l\in[1,N]$. The control excitation is introduced by a long single-photon pulse with a Gaussian profile, as defined in Eq. (\ref{eq:cin}).
\par Subsequently, the system evolves under the non-Hermitian Hamiltonian and in the presence of the input pulse is given by the equation
\begin{equation}
\frac{d}{dt}\ket{\Psi(t)}=-i \bold{H_{NH}} \ket{\Psi(t)}+i\sqrt{2d_\mathrm{c1}\gamma_{e_\mathrm{c}}}c_\mathrm{in}(t)\sum_{l=1}^N \ket{e_\mathrm{c}}_l.
\end{equation}
The norm of the system's wavefunction and the input pulse at time $t$ is
\begin{align}\notag
&\langle\Psi(t)|\Psi(t)\rangle+\int^{t_\mathrm{max}}_{t}dt\langle\Psi_\mathrm{in}(t)|\Psi_\mathrm{in}(t)\rangle=\\&=\sum^N_l|c_{e_\mathrm{c}}(t)|^2+\sum^N_l|c_{r_\mathrm{c}}(t)|^2+\int_t^{t_\mathrm{max}}|c_\mathrm{in}(t)|^2.
\end{align}
The value of the norm is unity at $t=t_0$ and is gradually reduced under the evolution of the non-Hermitian dynamics, while the input enters the dissipative system.

\par The process begins with a quantum jump, which always occurs since the incident control photon will either be subject to a dephasing quantum jump or the photon will leave the system, corresponding to a decay jump. The time of the jump is determined by the standard stochastic wfMC procedure. A value between 0 and 1 is randomly chosen and the time of the jump is set to the temporal point, when the norm reaches that value.
\par Once the time of the first jump $t_j$ is set, the nature of the jump is determined through a second stochastic process. It can be one of the four, described by the jump operators (\ref{eq:fsdjum}),(\ref{eq:fsgjum}),(\ref{eq:djump}), and (\ref{eq:ejump}). The non-normalized probabilities of these jumps are
\begin{align}
&p_{\gamma_{ge_\mathrm{c}}}(t_{j})=2\gamma_{e_\mathrm{c}}\sum_{k=1}^N\left|c_{e_\mathrm{c}}^k(t_{j})\right|^2, \label{eq:fspegc}
\\&p_{d_\mathrm{c}}(t_{j})=\left|i\sqrt{2d_\mathrm{c1}\gamma_{e_\mathrm{c}}}\sum_{k=1}^Nc^k_{e_\mathrm{c}}(t_{j})-c_\mathrm{in}(t_{j})\right|^2, \label{eq:pkc}
\\&p_{d_\mathrm{p}}(t_{j})=d_\mathrm{p1}\sum_{k=1}^N\left|\sum^N_{\substack{l=1\\ l\neq k}}\sqrt{\tilde{\gamma}_{ge_\mathrm{p}}^{k,l}}\right|^2|c^k_{r_\mathrm{c}}(t_{j})|^2, \label{eq:pkp}
\\&p_{\gamma_{ge_\mathrm{p}}}(t_{j})
=\sum_{k=1}^N\sum_{\substack{l=1\\l\neq k}}^N\left|\sqrt{\tilde{\gamma}_{ge_\mathrm{p}}^{k,l}}\right|^2|c^k_{r_\mathrm{c}}(t_{j})|^2,\label{eq:fspegp}
\end{align}
where the dephasing rate $\tilde{\gamma}_{ge_\mathrm{p}}^{k,l}$ was introduced in Eq. (\ref{eq:fserate}).
Furthermore, the normalized probabilities $\Pi_i$ for $i$=$\{\gamma_{ge_\mathrm{c}},d_\mathrm{c},d_\mathrm{p},\gamma_{ge_\mathrm{p}}\}$ are given by a normalization process as $\Pi_i=p_i(t_j)/\sum_ip_i(t_j)$.
Depending on the nature of the jump, the evolution of the system is determined. The possible outcomes are categorized as follows:
\par\textit{Decay jumps $\mathit{\hat{L}_{d_\mathrm{c}}}$, $\mathit{\hat{L}_{\gamma_{ge_\mathrm{c}}}}$.---} If either of the two decay jumps occurs, the incident control photon exits the system through spontaneous emission from the excited state or transmission through the atomic ensemble. This type of decay jump marks the end of the trajectory, and the simulation is concluded.

\par\textit{Dephasing jump $\mathit{\hat{L}_{\gamma_{d_\mathrm{p}}}}$.---} In the event of a dephasing jump due a change in transmittance of the probe, the control input photon is absorbed by the system, and the state of the system reads
\begin{equation}
\ket{\Psi'(t_{j})}=\frac{\hat{L}_{d_\mathrm{p}}\ket{\Psi(t_{j})}}{\sqrt{p_{\kappa_\mathrm{p}}(t_{j})}}.
\end{equation}
Subsequently, the process is repeated in order to determine the time and nature of the following jump. Since the incident input-photon pulse was absorbed, we set $c_\mathrm{in}(t)=0$ for $t>t_{j}$ and the system evolves according to the equation $\frac{d}{dt}\ket{\Psi'(t)}=-i \bold{H_{NH}} \ket{\Psi'(t)}$. 
\par\textit{Dephasing jump $\mathit{\hat{L}_{\gamma_{ge_\mathrm{p}}}}$.---} When a dephasing jump occurs due to the spontaneous emission of an excited atom in the probe branch, an additional stochastic process becomes essential to determine which specific atom among the ensemble of $N$ atoms decayed from the excited state $\ket{e_\mathrm{p}}$ of the probe branch. This process enables the identification of the atom responsible for the effective dephasing. The normalized probability of the $l$th atom to decay is given by Eq. (\ref{eq:pi}).
After identifying the atom $l$ that decayed, the input control photon is absorbed by the system, resulting in the preparation of the state described in Eq. (\ref{eq:psi}). This state exhibits localization around the $l$th atom that underwent decay, as explained in Sec. \ref{sec:fsprob}. The aforementioned process is then repeated to determine the timing and nature of the subsequent jump.
\par The process is repeated as many times as necessary until the control excitation undergoes decay via either of the two decay jumps ($\hat{L}_{ge_\mathrm{c}}$, $\hat{L}_{d_\mathrm{c}}$), or until three dephasing jumps ($\hat{L}_{d_\mathrm{p}}$) occur, which we again take as the threshold for a successful trajectory.
\par
\begin{figure}
\centering
  \includegraphics[width=1.\linewidth]{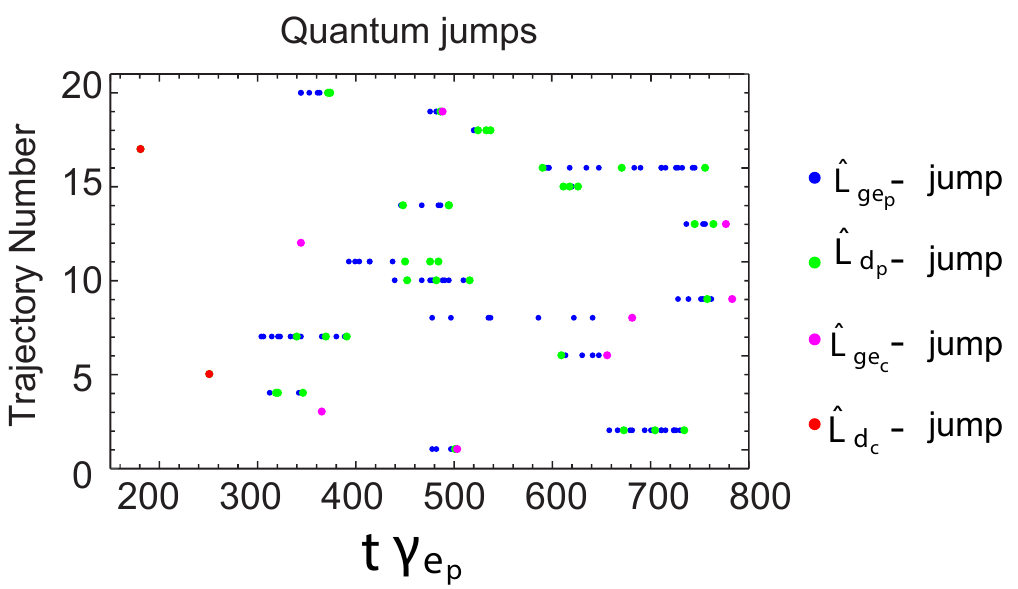}
    \caption{First 20 trajectories of the Monte Carlo simulation for 1D Gaussian distribution of $N=1000$ atoms in free space for optical depth $d_\mathrm{c}=100$ and blockaded optical depth $\overline{d}_\mathrm{b,p}=2$. The color of the dots indicates the nature of the jump that occurred at the specific time. The parameters are fixed to $\Delta/\gamma_{e_\mathrm{c}}=40, d_\mathrm{1p} = d_\mathrm{1c} = \gamma_{e_\mathrm{p}}/\gamma_{e_\mathrm{c}} = 1, \Omega_\mathrm{c}/\Delta=0.05, \Omega_\mathrm{p}/\gamma_{e_\mathrm{c}}=10, \delta/\gamma_{e_\mathrm{c}}=0.113, |\alpha_{\mathrm{in,p}}|^2/\gamma_{e_\mathrm{c}}=0.32$.}
    \label{fig:fsjum}
\end{figure}

 The numerical results are averaged over $N_\mathrm{traj}=2000$ trajectories for each value of the optical depth $d_\mathrm{c}$ of the control branch and the average blockaded optical depth $\overline{d}_\mathrm{b,p}=1/N\sum_k^Nd_\mathrm{b,p}^k$ of the probe branch. The 20 first trajectories of the $ d_\mathrm{c}=100$, $\overline{d}_\mathrm{b,p}=2$ simulation for an atomic ensemble with a 1D Gaussian distribution are presented in Fig. \ref{fig:fsjum}. It is evident that every trajectory ultimately is concluded either as successful, characterized by three $\hat{L}_{d_\mathrm{p}}$ dephasing jumps (green dots), or as unsuccessful, indicated by one of the decay jumps (pink and red dots). 
Moreover, it is noteworthy that the event of a $\hat{L}_{ge_\mathrm{p}}$ dephasing jump due to spontaneous emission of a probe photon is rarely succeeded by a $\hat{L}_{d_\mathrm{c}}$ decay jump, denoting the loss of the control photon by transmission through the ensemble. This is due to the fact that the localization effect induced by the dephasing jump strongly suppresses the propagation of the photon.

\subsection{\label{sec:resfs}Results}

The first part of this section focuses on the impedance-matching conditions, which are numerically optimized and compared with our analytical estimate derived in Sec. \ref{sec:fsim}. The second part of the section discusses the numerical optimization of the efficiency of the SPT versus the optical depth $d_\mathrm{c}$ of the control branch and the average blockaded optical depth $\overline{d}_\mathrm{b,c}$ of the probe branch.
\par\textit{Impedance matching.---} Similar to the cavity case, the impedance matching probability $P_\mathrm{{IM}}$ is defined as the ratio of the probability of the first jump being a dephasing jump ($\hat{L}_{{ge_\mathrm{p}}}$, $\hat{L}_{d_\mathrm{p}}$), instead of a decay jump ($\hat{L}_{ge_\mathrm{c}}$, $\hat{L}_{d_\mathrm{c}}$). This is given by the relation
\begin{equation}
P_{\mathrm{IM}}=\frac{\int_{t_0}^{t_{max}}dt\left(p_{\gamma_{d_\mathrm{p}}}(t)+p_{\gamma_{ge_\mathrm{p}}}(t)\right)}{\int_{t_0}^{t_{max}}dt\left(p_{\gamma_{ge_\mathrm{c}}}(t)+p_{d_\mathrm{c}}(t)+p_{\gamma_{d_\mathrm{p}}}(t)+p_{\gamma_{ge_\mathrm{p}}}(t)\right)},
\end{equation}
where the probabilities for each jump are defined in Eqs. (\ref{eq:fspegc})--(\ref{eq:fspegp}).
$P_{\mathrm{IM}}$ being close to unity plays a critical role in the operation of the SPT and serves as an upper limit for its efficiency, as it gives the probability of successful absorption of the control photon by the atomic ensemble.
\par
\begin{figure}[t!]
\includegraphics[width=.9\columnwidth]{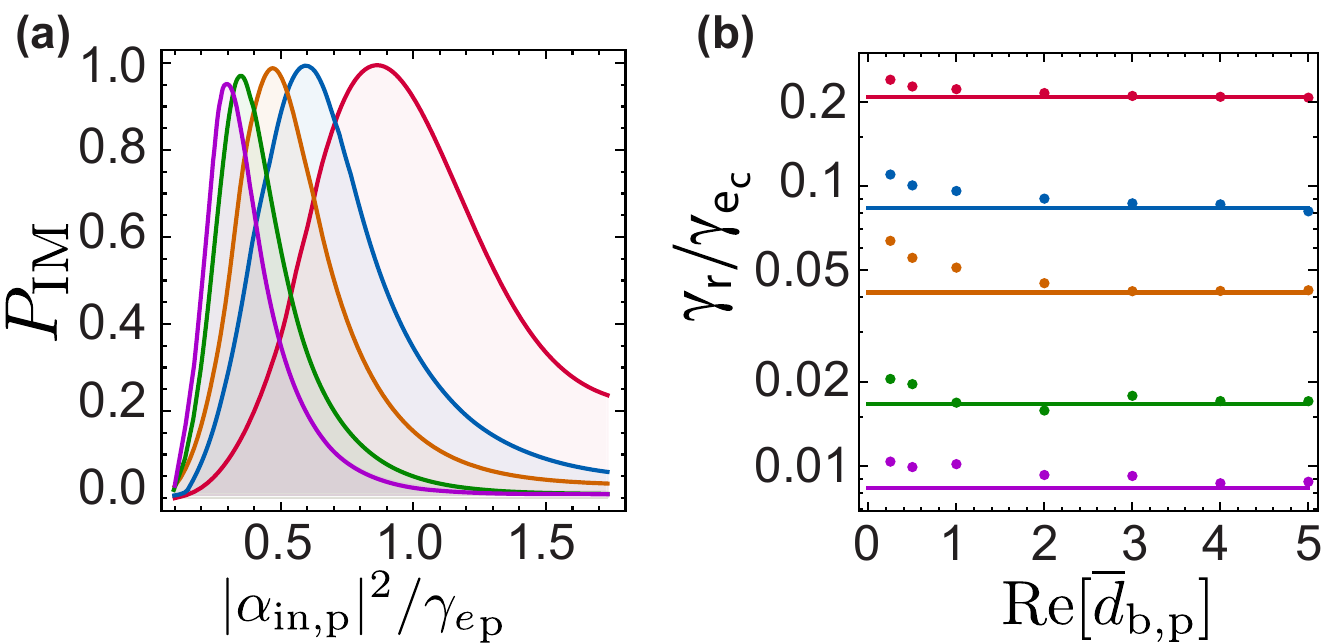}
\caption{Impedance-matching results for the free-space model with $N = 1000$ atoms distributed in a Gaussian profile along the $z$ axis.
(a) Impedance-matching probability $P_\mathrm{IM}$ as a function of probe strength for $\mathrm{Re}[\overline{d}_\mathrm{b,p}] = 2$, with multiple curves corresponding to different values of optical depth $d_\mathrm{c} = 20, 40, 100, 200, 500$ (lower curves correspond to lower $d_\mathrm{c}$).
(b) Numerically optimized probe-induced dephasing rate (dots) and theoretical estimate (solid line) as a function of the blockaded optical depth, shown for the same values of $d_\mathrm{c}$.
The parameters are fixed to $\Delta/\gamma_{e_\mathrm{c}} = 4d_\mathrm{c}$, $\Omega_\mathrm{c} = \Delta/40$, and $\Omega_\mathrm{p}/\gamma_{e_\mathrm{c}} = 10$, with $d_\mathrm{1p} = d_\mathrm{1c} = \gamma_{e_\mathrm{p}}/\gamma_{e_\mathrm{c}} = 1$. The detuning $\delta$ is optimized at each point to enhance $P_\mathrm{IM}$.}
\label{fig:IMfs}
\end{figure}

 In Fig. \ref{fig:IMfs}(a) we plot the impedance-matching probability versus the probe's strength. As seen in the figure, the $P_{\mathrm{IM}}$ can be optimized by varying the strength of the incident probe input field strength. In the optimization, the parameters are chosen such that the derived conditions for the detunings are also fulfilled. The simulations are repeated for different values of the control branch's optical depth, i.e., $d_\mathrm{c}=20,40,100,200,500$ and the probe branch's blockaded cooperativity, i.e., $\mathrm{Re}[\overline{d}_\mathrm{b,p}]\approx0.25,0.5,1,2,3,4,5$ for the 1D Gaussian atomic distribution. 
\par The value of $P_\mathrm{IM}$ observed is very close to the theoretically estimated value in Eq. (\ref{eq:imcfs}). The maximum value $P_{\mathrm{IM}}=0.992$ is obtained for $d_\mathrm{c}=500$, being close to the theoretical estimate which is $P_{\mathrm{IM}}=0.998$. The difference is attributed to the non-homogeneous nature of the Gaussian distribution.
\par Moreover, in Fig. \ref{fig:IMfs}(b) we plot the numerically optimized impedance matching conditions for the average dephasing rate $\gamma_r^\mathrm{opt}=\frac{1}{N}\sum_k^N(\gamma_{d_\mathrm{p}}^k+\sum_l^N\gamma_{e_\mathrm{p}}^{k,l})$ (dots) and the theoretical estimate (solid line) in Eq. (\ref{eq:gopt}), versus the blockaded optical depth. We find good agreement between the two, with only a slight deviation attributable to the system's inherent inhomogeneity. For lower values of $\overline{d}_\mathrm{b,p}$ and large $d_\mathrm{c}$ the deviation that is observed is due to a long plateau of almost constant $P_{\mathrm{IM}}$, where we have chosen the highest value.

\par\textit{Efficiency.---}As in the cavity case, the efficiency is defined as the number of trajectories in which at least three $\hat{L}_{d_\mathrm{p}}$ jumps occur, over the total number of trajectories
\begin{equation}
\eta=\frac{N_\mathrm{traj}(N_{\gamma_{d_\mathrm{p}}}^\mathrm{th}\geq3)}{N_\mathrm{traj}}.
\label{eq:effs}
\end{equation}
\par\textit{Gain.---}The gain $\mathcal{G}$ is defined as the number of probe-induced dephasing $\hat{L}_{d_\mathrm{p}}$ jumps occurring before the control excitation decays, averaged over all trajectories contributing to the efficiency.

\begin{figure}[t!]
\includegraphics[width=1.\columnwidth]{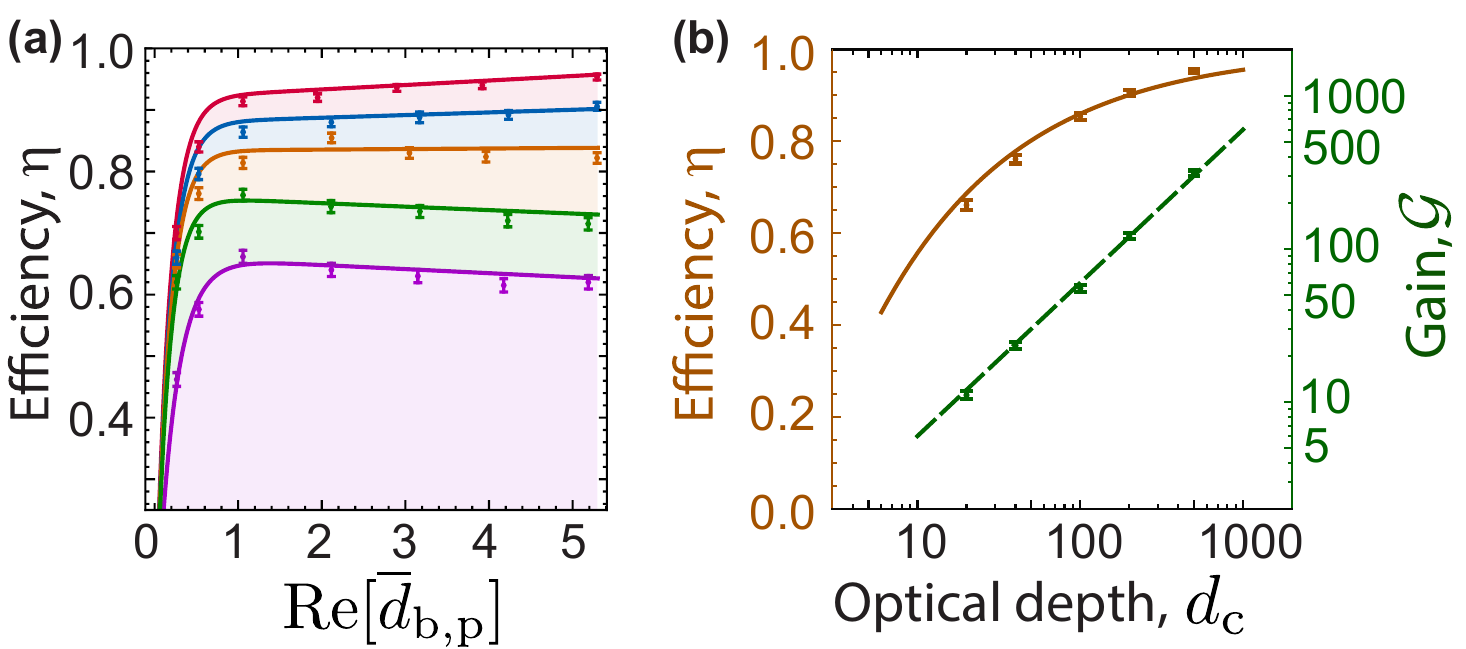}
\caption{Performance of the free-space model with $N = 1000$ atoms distributed in a Gaussian profile along the $z$ axis.
(a) SPT efficiency as a function of blockaded optical depth, for control branch optical depths $d_\mathrm{c} = 20, 40, 100, 200, 500$ (lower curves correspond to lower $d_\mathrm{c}$).
(b) SPT efficiency (solid) and gain (dashed) as functions of $d_\mathrm{c}$, with the blockaded optical depth fixed to its optimal value at each point.
The parameters are fixed to $\Delta/\gamma_{e_\mathrm{c}} = 4d_\mathrm{c}$, $\Omega_\mathrm{c} = \Delta/40$, and $\Omega_\mathrm{p}/\gamma_{e_\mathrm{c}} = 10$, with $d_\mathrm{1p} = d_\mathrm{1c} = \gamma_{e_\mathrm{p}}/\gamma_{e_\mathrm{c}} = 1$.
The detuning $\delta$ and probe strength $|\alpha_{\mathrm{in,p}}|^2$ are optimized at each point to enhance efficiency. Solid curves are polynomial fits provided as guides to the eye; points correspond to simulation data.}
\label{fig:effs}
\end{figure}

In Fig. \ref{fig:effs}, the results are depicted for different values of the control optical depth, i.e., $d_\mathrm{c}=20,40,100,200,500$ and the probe blockaded optical depth, i.e., $\mathrm{Re}[\overline{d}_\mathrm{b,p}]\approx0.25,0.5,1,2,3,4,5$ for 1D Gaussian atomic distribution. As shown Fig.~\ref{fig:effs}, the maximum efficiency is obtained for the highest considered value of the optical depth $d_\mathrm{c}=500$ and $\overline{d}_\mathrm{b,p}\approx5$, reaching $0.954$ with an associated gain of $\mathcal{G}=312$. 
Higher values of control optical depth are observed to result in increased efficiency and gain. Consequently, further performance improvements can be achieved by increasing the optical depth. The observed efficiency exceeds that of previous Rydberg-based SPTs by more than 15\%~\cite{Gorniaczyk2016} and is achieved at an optical depth that has been demonstrated experimentally~\cite{Sprakes2013}. Relevant physical values and experimental parameters are discussed in Appendix \ref{appendix:b}.

\section{\label{sec:concl}Conclusions}
We have conducted a comprehensive analysis and characterization of different variants of all-optical SPTs capable of operating in the cw regime. These SPTs are based on ensembles of Rydberg atoms and span both free space and cavity configurations with varying geometries. We analyzed the optimal impedance-matching conditions required for the efficient capture of a single photon, which control the transmission and reflection of the probe field. Additionally, we utilized engineered probe-induced dephasing to optimize the overall efficiency and gain of the devices. The estimated efficiencies reach up to 95.4\% with a corresponding SPT gain of up to 312, although a full optimization over all parameters has not been performed and further improvement may be possible. The estimated efficiencies exceed previous Rydberg-based SPT realizations by more than 15\%, with a gain more than three times higher~\cite{Gorniaczyk2016}, using experimentally demonstrated parameters~\cite{Sprakes2013,Sauer2004}. While our numerical simulations were limited by computational constraints, we expect the efficiency and gain to further increase with larger ensemble sizes. The proposed devices may expand the frontiers of cw control at the single-photon level, leading to the development of novel tools for optical control and quantum information processing.

\section*{Aknowledgments}
We thank Jens Eisert for valuable discussions. This work was supported by the ERC grant QIOS (Grant No. 306576). I.T. acknowledges funding by the Deutsche Forschungsgemeinschaft through the Emmy Noether program (Grant No. ME 4863/1-1). A.S. acknowledges the support of Danmarks Grundforskningsfond (DNRF 139, Hy-Q Center for Hybrid Quantum Networks). O.K. acknowledges the support from EPSRC grants EP/V00171X/1 and EP/X017222/1, and NATO SPS project MYP.G5860.
\\\\
\renewcommand{\appendixname}{APPENDIX}
\appendix
\section{INTERATOMIC RYDBERG-RYDBERG INTERACTION}
\label{appendix:a}

In this appendix, we examine the interaction between atoms excited to Rydberg states and analyze its dependence on the principal quantum numbers of the states. This analysis serves to justify the assumption of neglecting same-branch interactions between atoms in state $\ket{r_\mathrm{p}}$, as stated in Secs. \ref{sec:modcav} and \ref{sec:sysfs}.

When considering two atoms in Rydberg states $\ket{r}$ and $\ket{r'}$, characterized by principal quantum numbers $n$ and $n'$ respectively, the primary electrostatic interaction between them is the dipole-dipole interaction. This interaction predominates at interatomic distances exceeding their respective radii, i.e., $R/a_0 \gg \max( n^2, n'^2)$, where $R$ is the interatomic distance and $a_0$ the Bohr radius. The dipole-dipole interaction is described by the operator
\begin{equation}
\hat{V}_{dd} = \frac{1}{4\pi\epsilon_0} \frac{\mathbf{\hat{d}_1} \cdot \mathbf{\hat{d}_2} - 3(\mathbf{\hat{d}_1} \cdot \mathbf{e_R})(\mathbf{\hat{d}_2} \cdot \mathbf{e_R})}{R^3},\label{eq:ddop}
\end{equation}
where $\epsilon_0$ is the vacuum permittivity, $\mathbf{e_R}$ is the unit vector along the relative coordinate between the two atoms, and $\mathbf{\hat{d}_1}$ and $\mathbf{\hat{d}_2}$ are the electric dipole operators of the first and second atom, respectively.

The dipole-dipole interaction induces transitions between the initial two-atom state $\ket{r}\ket{r'}$ and other two-atom states, governed by the conventional dipole selection rules. However, in practice, this interaction predominantly couples a limited number of the closest two-atom states satisfying these selection rules to the initial state, which have small energy differences and large dipole matrix elements. To a good approximation, we can consider the long-range interaction between Rydberg atoms as predominantly arising from coupling to a neighboring two-atom state, i.e., $\ket{r}\ket{r'}\leftrightarrow\ket{r_a}\ket{r_b}$, where the principal quantum numbers of states $\ket{r_a}$ and $\ket{r_b}$ are $n_a = n+1$ and $n_b = n'-1$ respectively. The
energy difference between the two two-atom states is given by the Förster defect $\delta_F = E_r + E_{r'} - E_{r_a} - E_{r_b}$, where $E_i$ denotes the energy of state $\ket{i}$.

The time-independent Schr\"odinger equation describing the dipole interaction between the two-atom states in matrix form reads
\begin{equation}
\left(
\begin{array}{cc}
 \delta_F & V_{dd} \\
V_{dd}^\dagger & 0 \\
\end{array}
\right)\left(\begin{array}{cc}
\ket{r}\ket{r'}
\\ \ket{r_a}\ket{r_b}
\end{array}\right)=U\left(\begin{array}{cc}
\ket{r}\ket{r'}
\\ \ket{r_a}\ket{r_b}
\end{array}\right).
\end{equation}
Solving for $\ket{r_a}\ket{r_b}$ and substituting into the second row leads to an eigenvalue equation for $\ket{r}\ket{r'}$
\begin{equation}
\frac{\hat{V}_{dd}^\dagger\hat{V}_{dd}}{U-\delta_F}\ket{r}\ket{r'}=U\ket{r}\ket{r'}.
\label{eq:eigeneq}\end{equation}
Using Eq. (\ref{eq:ddop}) we have that 
\begin{align}
    \bra{r_ar_b}V_{dd}^\dagger V_{dd}\ket{rr'}&=\frac{(\bra{r_a} \hat{d}_1\ket{r})^2(\bra{r_b} \hat{d}_2\ket{r'})^2}{R^6}\nonumber
    \\&\approx\frac{e^2n^4n'^4a_0^2D_\phi}{R^6},
\label{eq:vv}\end{align}
where is $e$ the electron charge and $D_\phi$ a coefficient arising from the angular part of the dipole operators \cite{Saffman2008}.
Using Eq. (\ref{eq:vv}), we can solve Eq. (\ref{eq:eigeneq}), and the interaction energy is found to be
\begin{equation}
    U_{rr'}(R)\approx\frac{\delta_F}{2}-\mathrm{sgn}(\delta_F)\sqrt{\frac{\delta_F^2}{4}+\frac{e^2n^4n'^4a_0^2D_\phi}{R^6}},
\end{equation}
which, for large interatomic distances, takes the van der Waals form
\begin{equation}
    U_{rr'}(R)\approx \frac{e^2n^4n'^4a_0^2D_\phi}{\delta_F R^6}.
\end{equation}
Expressing the interaction energy as $ U_{rr'}(R)=C_{6,rr'}/R^6$, and using the fact that the energy difference between neighboring Rydberg states scales with the inverse of the principal quantum number to the power of 3, i.e.,  $E_r-E_a\propto n^{-3}$ and $E_{r'}-E_b\propto n'^{-3}$, leading to $\delta_F\propto \min(n,n')^{-3}$, we deduce that the relation for the interaction coefficient scales as

\begin{figure}[t!]
\includegraphics[width=1.\columnwidth]{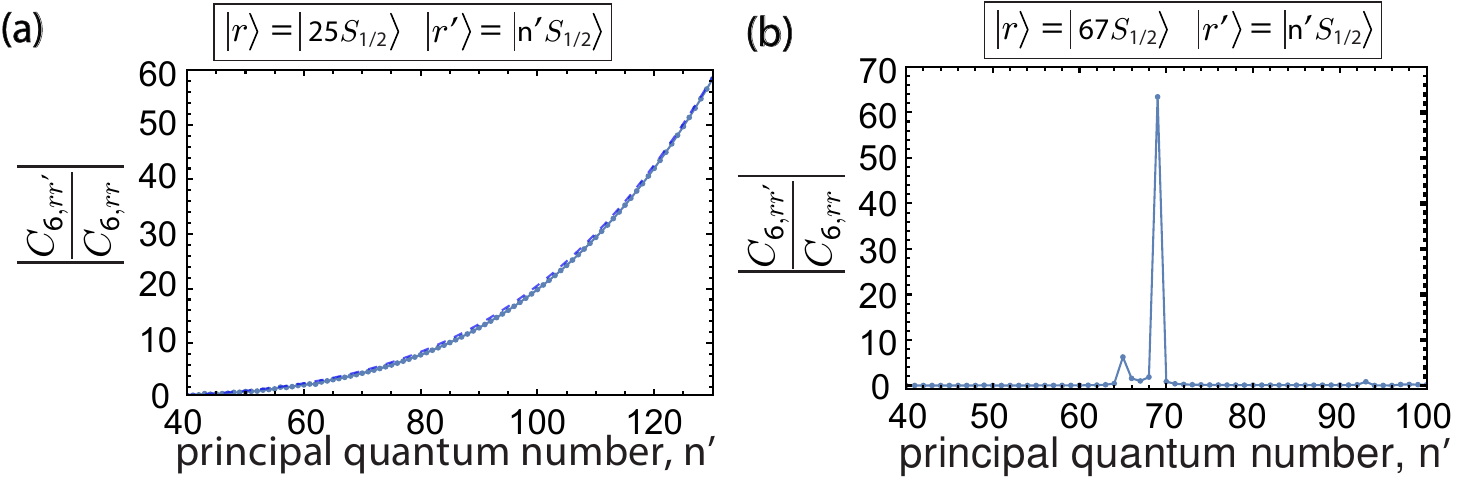}
\caption{(a) Absolute value of the ratio of the interaction coefficient between $^{87}$Rb atoms in state $\ket{25S_{1/2}}$ and in state $\ket{n'S_{1/2}}$ over the interaction coefficient between $^{87}$Rb atoms in state $\ket{25S_{1/2}}$, as a function of the principal quantum number $n'$ (solid line). The function $(n'/25)^4/13$ is shown as a dashed line. (b)Absolute value of the ratio of the interaction coefficient between $^{87}$Rb atoms in state $\ket{67S_{1/2}}$ and in state $\ket{n'S_{1/2}}$ over the interaction coefficient between $^{87}$Rb atoms in state $\ket{67S_{1/2}}$, as a function of the principal quantum number $n'$.}
\label{fig:app1}
\end{figure}

\begin{equation}
    C_{6,rr'}\propto n^4n'^4\min(n,n')^3.
\end{equation}
It follows that the interaction coefficient between two atoms in state $\ket{r}$, characterized by the principal quantum number $n$, reads $C_{6,rr}\propto n^{11}$. Consequently, angular parts of the wave-functions, the ratio of the interaction coefficient between two atoms in states  $\ket{r}$ and  $\ket{r'}$ relative to that between two atoms both in state  $\ket{r}$, for $n'>n$, obeys the scaling relation
\begin{equation}
    \frac{|C_{6,rr'}|}{|C_{6,rr}|}\propto (n'/n)^4.
\end{equation}
As a result, when the principal quantum number of state $\ket{r'}$ greatly exceeds that of $\ket{r}$, the interaction between atoms in states $\ket{r}$ becomes much weaker compared with the interaction between atoms in states $\ket{r}$ and $\ket{r'}$, rendering it negligible.

We illustrate this in Fig. \ref{fig:app1}(a) by plotting the absolute value of the ratio of the interaction coefficient between $^{87}$Rb atoms in state $\ket{25S_{1/2}}$ and in state $\ket{n'S_{1/2}}$ over the interaction coefficient between $^{87}$Rb atoms in state $\ket{25S_{1/2}}$, versus the principal quantum number $n'$. These interaction coefficients were calculated using the Alkali Rydberg Calculator (ARC) python package \cite{ARC}, assuming that the angular parts of the wave-functions are identical. By fitting the calculated results, we find that $\left|\frac{C_{6,25S_{1/2}n'S_{1/2}}}{C_{6,25S_{1/2}25S_{1/2}}}\right| \approx \left(\frac{n'}{25}\right)^4/13$ for $40 \leq n' \leq 130$. We note that we are interested in the absolute value of the interaction coefficient $C_{6,rr'}$, as the sign does not affect the interaction's purpose of shifting the probe branch out of EIT resonance. This holds true regardless of whether the resonance is shifted upward or downward in energy. Consequently, the Rydberg-Rydberg interaction can be either attractive or repulsive.

We have demonstrated that interaction between atoms in state $\ket{r}$ can generally be neglected compared with the interaction between atoms in states $\ket{r}$ and $\ket{r'}$ when $n'\gg n$. However, we can relax this condition for special values of $n$ and $n'$, when states $\ket{r}$ and $\ket{r'}$ are close to a Förster resonance together with two neighboring states. A Förster resonance has a nearly zero Förster defect $\delta_F$ among the four states, resulting in a very large interaction coefficient $C_{6,rr'}$. The significant disparity between $C_{6,rr'}$ and $C_{6,rr}$ again allows us to neglect the interaction between atoms in states $\ket{r}$, this time without the precondition of $n'\gg n$. 

We demonstrate the effect of a Förster resonance in Fig. \ref{fig:app1}(b), where we plot the absolute value of the ratio of the interaction coefficient between $^{87}\mathrm{Rb}$ atoms in state $\ket{67S_{1/2}}$ and state $\ket{n'S_{1/2}}$ over the interaction coefficient between $^{87}\mathrm{Rb}$ atoms in state $\ket{67S_{1/2}}$, as a function of the principal quantum number $n'$. We observe a Förster resonance occurring at $n'=69$ between $^{87}\mathrm{Rb}$ atoms in states $\ket{67S_{1/2}}$ and $\ket{69S_{1/2}}$, resulting in a much larger interaction coefficient compared with other values of $n'$. We note that this Förster resonance was utilized in Ref. \cite{Tiarks2014}.

In connection with Secs. \ref{sec:modcav} and \ref{sec:sysfs}, we designate the Rydberg state $\ket{r'}$ as $\ket{r_\mathrm{c}}$ and $\ket{r}$ as $\ket{r_\mathrm{p}}$. Consequently, when the principal quantum number of state $\ket{r_\mathrm{c}}$ significantly exceeds that of $\ket{r_\mathrm{p}}$ or alternatively when $\ket{r_\mathrm{c}}$ and $\ket{r_\mathrm{p}}$ are close to a Förster resonance, the interaction between atoms in state $\ket{r_\mathrm{p}}$ becomes negligible compared with that between atoms in states $\ket{r_\mathrm{p}}$ and $\ket{r_\mathrm{c}}$. It's important to note that since we consider a single photon in the control branch, interactions between atoms in states $\ket{r_\mathrm{c}}$ are absent. Therefore, we exclusively consider interactions between atoms in Rydberg states from different branches, i.e., between atoms in states $\ket{r_\mathrm{p}}$ and $\ket{r_\mathrm{c}}$.

\section{EXPERIMENTAL CONSIDERATIONS} \label{appendix:b}
In this appendix, we consider the relative physical values and experimental parameters for the two schemes of the proposed device, as described in Secs. \ref{sec:cav} and \ref{sec:fs}, inspired by similar experimental setups.

Initially we focus on the cavity scheme, following the experimental setup of Ref. \cite{Vaneecloo2022}. An atomic ensemble of 800 $^{87}$Rb atoms with relevant level structure described in Sec. \ref{sec:modcav} is placed inside a cavity with a decay rate of $\kappa_\mathrm{p}\approx\kappa_\mathrm{c}=2\pi\cross 2.9$ MHz. The ensemble has a Gaussian root mean square radius $\sigma=5$ $\mathrm{\mu}$m in all three dimensions. The ground state of $^{87}$Rb is $\ket{g}=\ket{5S_{1/2}}$. For excited states of the probe and control branch, we use states $\ket{e_\mathrm{p}}=\ket{5P_{1/2}}$ and $\ket{e_\mathrm{c}}=\ket{5P_{3/2}}$, respectively, with radiative decays of $\gamma_{e_\mathrm{p}}\approx\gamma_{e_\mathrm{c}}\approx2\pi\cross 3$ MHz. The coupling constants between ground and excited states are $g_\mathrm{c}\approx g_\mathrm{p}=2\pi\cross10$ MHz. Subsequently, the cooperativities of control and probe branches read $C_\mathrm{p}\approx C_\mathrm{c}=g_\mathrm{c}^2/\kappa_\mathrm{c}\gamma_{e_\mathrm{c}}=11.5$. 

\begin{figure}[t!]
\includegraphics[width=1.\columnwidth]{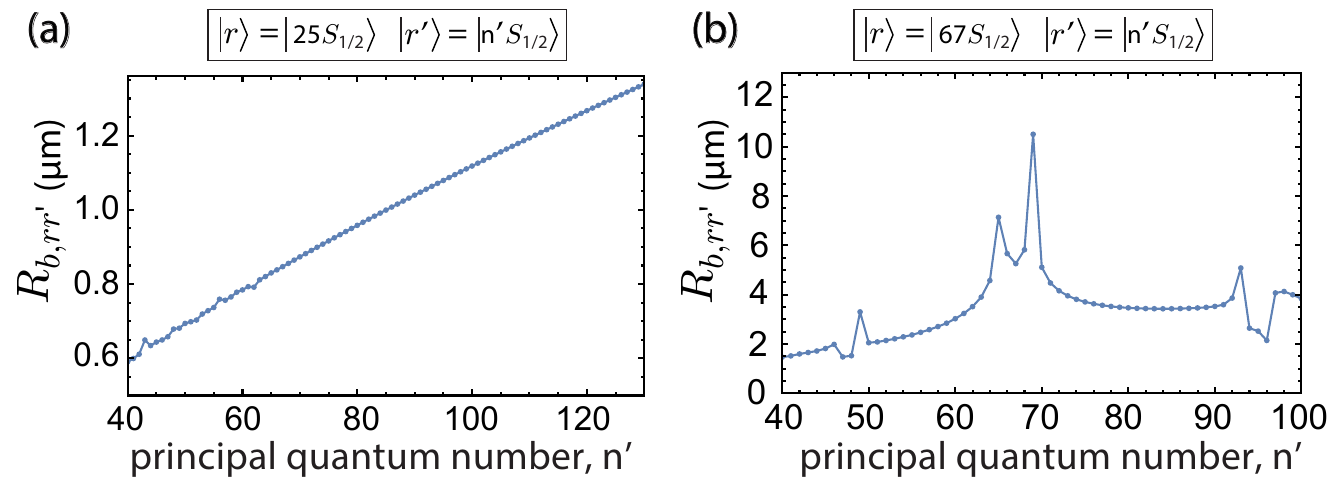}
\caption{The blockaded radius due to interaction between $^{87}$Rb atoms in (a) state $\ket{25S_{1/2}}$ and state $\ket{n'S_{1/2}}$, (b) state $\ket{67S_{1/2}}$ and state $\ket{n'S_{1/2}}$, as a function of the principal quantum number $n'$. $\Omega_\mathrm{p}/\gamma_{e_\mathrm{p}}=10$, and $\gamma_{e_\mathrm{p}}=2\pi\cross3$ MHz.}
\label{fig:app2}
\end{figure}

The blockaded cooperativity of the probe branch can be approximated as $C_\mathrm{b,p}\approx C_\mathrm{p} (R_{\mathrm{b},rr'}/\sigma)^3$, where $R_{\mathrm{b},rr'}$ is the blockaded radius due to interaction between atoms in Rydberg states $\ket{r}$ and $\ket{r'}$ defined as $R_{\mathrm{b},rr'}=(C_{6,rr'}\gamma_{e_\mathrm{p}}/|\Omega_\mathrm{p}|^2)^{1/6}$. In Fig.~\ref{fig:app2}(a), we plot the blockaded radius due to interaction between atoms in states $\ket{25S_{1/2}}$ and $\ket{n'S_{1/2}}$ as a function of the principal quantum number $n'$. 

Considering the following probe and control Rydberg states, $\ket{r_\mathrm{p}} = \ket{25S_{1/2}}$ and $\ket{r_\mathrm{c}} = \ket{125S_{1/2}}$, results in a blockaded radius of $R_{\mathrm{b},rr'} \approx 1.3 \ \mu\text{m}$, as shown in Fig.~\ref{fig:app2}(a). This corresponds to a blockaded cooperativity of $C_\mathrm{b,p} \approx 0.2$ and according to the results presented in Fig.~\ref{fig:jk}(c), for $C_\mathrm{c}=10$, the device efficiency is expected to be $\approx 20\%$. Increasing the number of $^{87}$Rb atoms in the cavity to 7000 or decreasing the cavity decay rate to $\kappa_\mathrm{c}=2\pi \times 0.33 \ \text{MHz}$ would raise the cooperativity to $C_\mathrm{c}\approx C_\mathrm{p} \approx 100$. Under these conditions, and choosing probe and control Rydberg states $\ket{r_\mathrm{p}} = \ket{25S_{1/2}}$ and $\ket{r_\mathrm{c}} = \ket{85S_{1/2}}$, would yield a blockaded radius $R_{\mathrm{b},rr'} \approx 1 \ \mu\text{m}$ corresponding to a blockaded cooperativity $C_\mathrm{b,p} \approx 0.8$, which would result in an efficiency of $\approx 55\%$ [see Fig.~\ref{fig:jk}(c)]. Further enlargement of the ensemble size or improvement of the cavity quality factor is projected to further enhance the device efficiency.

Furthermore, the radiative decay rates of the states $\ket{25S_{1/2}}$, $\ket{85S_{1/2}}$, and $\ket{125S_{1/2}}$ are $2\pi \times 6.1$ kHz, $2\pi \times 0.12$ kHz, and $2\pi \times 0.04$ kHz, respectively. These rates are much smaller than the decay rates of other processes in the system. Additionally, the maximum observed full protocol time is of the order of $t_{\mathrm{max}} = 10^3/\gamma_{e_\mathrm{p}} \approx 53$ $\mu$s, as shown in Fig.~\ref{fig:cavjum}. Moreover, the lifetimes of the Rydberg states $\ket{85S_{1/2}}$ and $\ket{125S_{1/2}}$  where the control excitation is stored are $0.7$ ms and $2.3$ ms, respectively, both of which exceed $t_\mathrm{max}$ by orders of magnitude. Therefore, decay of Rydberg states $\ket{r_\mathrm{p}}$, $\ket{r_\mathrm{c}}$ can be considered negligible during the protocol.

We then transition to the free space scheme, adopting parameters from Ref. \cite{Gorniaczyk2014}. In this setup, we consider an ensemble of $2.5 \times 10^4$ $^{87}$Rb atoms, initially prepared in their ground state $\ket{g} = \ket{5S_{1/2}}$. The ensemble has a Gaussian root mean square radius of $\sigma_\mathrm{fs} = 40$ $\mathrm{\mu m}$ along the $z$ axis. For the excited states of the probe and control branches, we again use $\ket{e_\mathrm{p}} = \ket{5P_{1/2}}$ and $\ket{e_\mathrm{c}} = \ket{5P_{3/2}}$, respectively, both having radiative decay rates of $\gamma_{e_\mathrm{p}} \approx \gamma_{e_\mathrm{c}} \approx 2\pi \times 3$ MHz.

The optical depths for the control and probe branches are $d_\mathrm{c} \approx d_\mathrm{p} = 25$. We approximate the blockade optical depth as $d_\mathrm{b,p} \approx d_\mathrm{p} R_{\mathrm{b},rr'} /  \sigma_\mathrm{fs}$. In Fig.~\ref{fig:app2}(b), we plot the blockaded radius due to interaction between atoms in states $\ket{67S_{1/2}}$ and $\ket{n'S_{1/2}}$ as a function of the principal quantum number $n'$. By choosing the states corresponding to the Förster resonance, $\ket{r_\mathrm{p}} = \ket{67S_{1/2}}$ and $\ket{r_\mathrm{c}} = \ket{69S_{1/2}}$, as the probe and control Rydberg states respectively, we achieve a blockade radius $R_{\mathrm{b},rr'} \approx 10.6$ $\mu$m and subsequently a blockaded optical depth $d_\mathrm{b,p}$ of 6.6. Based on this set of parameters, and the results presented in Fig.~\ref{fig:effs}(b), the device efficiency is estimated to be approximately $70\%$. If the number of $^{87}$Rb atoms in the ensemble was increased to $10^5$, the efficiency would rise to $\approx 85\%$, with further enhancement projected as the ensemble size grows.

We also note that the radiative decay rates for the states $\ket{67S_{1/2}}$ and $\ket{69S_{1/2}}$ are $2\pi \times 0.24$ kHz and $2\pi \times 0.22$ kHz, respectively, which are significantly lower than the decay rates associated with other processes in the system. Furthermore, the Rydberg state $\ket{69S_{1/2}}$, which stores the control excitation, has a lifetime of $0.36$ ms, far exceeding the maximum observed full protocol duration of approximately $40$ $\mu$s (see Fig.~\ref{fig:fsjum}).  Thus, the Rydberg states $\ket{r_\mathrm{p}}$ and $\ket{r_\mathrm{c}}$ can be considered long lived throughout the protocol.

\end{document}